\documentclass[10pt,twocolumn,twoside]{IEEEtran}
\usepackage{import}
\usepackage{color}
\usepackage{comment}
\usepackage{tikz}
\usetikzlibrary{arrows.meta}
\usepackage{graphicx}
\graphicspath{{}}
\newcommand{\com}[1]{{\color{black}{#1}}}
\newcommand{\fa}[1]{{\color{black}{#1}}}

\usepackage{amsfonts,amsmath,amsthm}
\usepackage{mathtools}
\usepackage{graphicx}
\usepackage{subcaption}
\usepackage{array}
\usepackage{tikz}
\usepackage{color,soul}
\usepackage{cmap}\usepackage[T1]{fontenc}

\soulregister\cite7
\soulregister\ref7
\soulregister\pageref7
\usetikzlibrary{arrows, shapes.geometric, patterns, decorations.markings}

\newtheorem{assumption}{Assumption}
\newtheorem{definition}{Definition}
\newtheorem{theorem}{Theorem}
\newtheorem{observation}{Observation}
\newtheorem{proposition}{Proposition}
\newtheorem{corollary}{Corollary}
\newtheorem{example}{Example}
\newtheorem{remark}{Remark}
\newtheorem{lemma}{Lemma}

\newcommand\numberthis{\addtocounter{equation}{1}\tag{\theequation}}
\newcommand{\natn}{\mathbb{N}_0}
\newcommand{\kinN}{k\in\natn}

\ifCLASSINFOpdf
\else
\fi
\begin{document}
	%
	% paper title
	% Titles are generally capitalized except for words such as a, an, and, as,
	% at, but, by, for, in, nor, of, on, or, the, to and up, which are usually
	% not capitalized unless they are the first or last word of the title.
	% Linebreaks \\ can be used within to get better formatting as desired.
	% Do not put math or special symbols in the title.
	\title{
		Max-Consensus {Over} Fading Wireless Channels}
	%
	%
	% author names and IEEE memberships
	% note positions of commas and nonbreaking spaces ( ~ ) LaTeX will not break
	% a structure at a ~ so this keeps an author's name from being broken across
	% two lines.
	% use \thanks{} to gain access to the first footnote area
	% a separate \thanks must be used for each paragraph as LaTeX2e's \thanks
	% was not built to handle multiple paragraphs
	%	
	\author{Fabio Molinari,
		Navneet~Agrawal,
		S\l awomir~Sta\'nczak,
		and~J\"org~Raisch% <-this % stops a space
		\thanks{F. Molinari is with the Control Systems Group - Technische Universit\"at Berlin, Germany. 
			(email: molinari@tu-berlin.de)}% <-this % stops a space
		\thanks{{N. Agrawal is with the Network Information Theory Group - Technische Universit\"at Berlin, Germany. 
			(email: navneet.agrawal@tu-berlin.de)}}% <-this % stops a space
		\thanks{{S. Sta\'nczak is with the Network Information Theory Group - Technische Universit\"at Berlin, Germany \& Fraunhofer Heinrich Hertz Institute, Germany.
				(e-mail: slawomir.stanczak@hhi.fraunhofer.de)}}% <-this % stops a space
		\thanks{J. Raisch is with the Control Systems Group - Technische Universit\"at Berlin, Germany
			(e-mail: raisch@control.tu-berlin.de)}% <-this % stops a space
		\thanks{This work was funded by the German Research Foundation (DFG) within their priority programme
			SPP1914 "Cyber-Physical Networking", \newline RA516/12-1 \&
			STA864/10-1}
		}
	\maketitle
	\IEEEpeerreviewmaketitle
% make the title area
\maketitle
% As a general rule, do not put math, special symbols or citations
% in the abstract or keywords.
\begin{abstract}
	The topic of this paper is achieving
	finite-time max-consensus
	%	within a finite number of iterations
	in a multi-agent system
	that communicates over a fading wireless
	channel \fa{and exploits its
		interference property}.
	\fa{This phenomenon corrupts the desired information
		when data is transmitted synchronously}.
	\fa{In fact, }%
	\fa{each transmitted signal is attenuated
		by an unknown and time-varying
		factor (fading coefficient), then,
		by interference,
		all such attenuated signals are
		summed up at a receiver}.
	\fa{Rather than
		combatting interference,
		we design a communication
		system that exploits it.
	Our strategy yields a more efficient
	usage of wireless resources
	compared to other algorithms.}
%	Compared to \com{other} algorithms,
%	the approach presented here
%	achieves max-consensus with a more
%	efficient usage of wireless resources.
%	In particular,
%	we design a multi-agent communication system
%	that exploits the superposition
%	property of the Wireless Multiple Access Channel
%	(WMAC). % to reach consensus efficiently.
	By simultaneously accessing
	this communication system,
	each agent obtains a weighted average
	of the neighbouring agents' information states.
%	Fewer wireless resources are required 
%	than with traditional approaches.
	With this piece of information at hand
	and with a switching consensus protocol employing
	broadcast authorisations for agents,
	max-consensus can be achieved
	within a finite number of iterations.
%	Starting from this, a switching 
%	max-consensus
%	protocol
%	for 
%	achieving finite-time max-consensus is obtained.
%	Randomized simulations prove that,
%	for achieving max-consensus,
%	employing the approach of this paper
%	requires fewer 
%	resources than by using traditional solutions
%	presented in literature.
%This paper presents
%a strategy
%for achieving max-consensus
%within a finite number
%of iterations
%in multi-agent systems
%communicating
%over fading wireless channels.
%A communication
%system,
%able 
%to harness the
%superposition property 
%of the wireless
%multiple access channel (WMAC),
%is designed.
%%\hl{The channel, according 
%%to the widely used 
%%  MAC\footnote{Multiple Access Channel} model,
%%is affected by 
%%real-valued fading coefficients}.
%Exploiting the superposition
%requires
%constructing, initially,
%a consensus protocol
%pledging 
%asymptotic
%convergence to
%max-consensus.
%This protocol employs
%a strategy based
%on a local broadcast 
%authorization
%for agents.
%%This strategy allows the system
%%for getting asymptotically 
%%to max-consensus.
%Then,
%a switching consensus protocol
%extending this
%idea of a local authorization for 
%agents
%is also presented.
%Employing this protocol
%guarantees a deterministic
%finite-time
%convergence to the max
%consensus
%and
%results
%%both in
%%a better convergence rate
%%and 
%in a better usage of 
%wireless resources
%compared to the
%traditional approaches.
\end{abstract}
%\import{./}{keywords.tex}

% Fabio
\section{Introduction}
Consensus is a useful 
notion 
in situations
where several autonomous intercommunicating
agents need to reach an agreement
over a variable of common interest,
see~\cite{ren2007information}.
Each agent has an estimate of this variable,
referred to as \textit{information state},
which is iteratively updated by
running a consensus protocol.
This is composed of two subsequent \com{steps}:
\textit{communication} and \textit{computation}.
In fact, first,
agents share their information states
with \com{their} neighbors.
Subsequently,
each agent uses
the received pieces of
information
to
update, at every iteration,
its information state.

When the system
aims at agreeing on the
average of initial information states,
we talk about \textit{average consensus}.
This has been extensively investigated in literature,
see, e.g.,~\cite{olfati2007consensus,cao2008reaching}.
Average consensus is employed in many
contexts, e.g., formation control 
of autonomous vehicles (see~\cite{ren2007consensus}),
flocking (see~\cite{olfati2004flocking}),
or
distributed solution of linear 
algebraic equations (see~\cite{mou2015distributed}).
The case of agents
seeking for an agreement on the 
largest information state,
namely \textit{max-consensus},
is also studied in literature, see, e.g.,~\cite{nejad2009max}.
Its applications range from
task allocation (see~\cite{brunet2008consensus})
to
road traffic automation
(see~\cite{molinari2018traffic}),
posture control (see~\cite{lippi2019distributed}),
and
distributed estimation (see~\cite{di2015distributed}).

Especially \com{in the case when}
agents are powered \com{by} batteries,
using resources economically is very important.
\cite{zheng2012fast} and \cite{goldenbaum2012nomographic}
introduce
a strategy for achieving consensus more efficiently
\com{in a wireless communication scenario}.
\com{This is achieved}
by exploiting \com{the interference}
property of the wireless channel,
which is
traditionally combatted.
In fact, 
in \com{standard} wireless settings,
interference is a phenomenon that 
needs to be eliminated.
However, as \com{shown} in \cite{goldenbaum2013harnessing},
creating an interference-free communication medium
for computing functions over the channel
is a suboptimal strategy.
Therefore, the design of consensus protocols
\com{aiming at exploiting} interference and, 
\com{as a consequence},
\com{allowing a more efficient use of} wireless resources,
has recently attracted attention.
\fa{\cite{molinari2018exploiting} presents
a protocol that achieves average consensus
and exploits the interference property of 
the fading wireless channel. The same is used
in \cite{molinari2019efficient}
for letting agents with single integrator dynamics
achieve a desired formation in space.
\cite{molinari2020exploiting} has introduced a 
consensus-based
privacy preserving strategy that exploits
interference for distributively
solving linear algebraic equations.}

To the best of our knowledge,
there \com{have been very few}
attempts 
to exploit interference
for achieving max-consensus.
Under the assumption
of an ideal channel,
\cite{iutzeler2012analysis} 
proposes a consensus protocol
that achieves max-consensus
with probability one,
by employing a random broadcast
strategy.
%Approaches using nomographic approximations
%of the max-function, see, e.g.,~\cite{goldenbaum2013harnessing},
%lead to approximation errors causing
%a diverging behavior.
%On the other hand, 
\cite{molinari2018exploitingMax}
suggests a switching
consensus protocol 
which exploits interference 
and is proven to achieve
max-consensus
deterministically.
%by exploiting the interference
%%and
%Unlike \cite{iutzeler2012analysis},
%\cite{molinari2018exploitingMax}
%is proven to work
%also when agents do not have the global
%knowledge of the network topology.
However, these attempts
do not deal with the presence of a fading
channel, \com{i.e., a channel that 
	attenuates each signal
by an unknown factor,}
which is required for a more realistic analysis.
The novelty of this paper
is a consensus protocol 
\com{which (i) achieves max-consensus in a 
	finite number of iterations,
	(ii) exploits interference,
	and (iii) can cope with
	fading channel coefficients}.

The paper is structured as follows:
%a protocol
%that circumvents this
%nonideality is presented.
in Section \ref{sec:probDesc},
\com{necessary facts from consensus and
	communication theory are collected}.
%the problem is presented
%under both control and communication
%points of view.
%Section \ref{sec:commsys}
%proposes a communication
%system that deals with
%the presence of a fading channel.
Section~\ref{sec:asyCons} 
presents a max-consensus protocol
\com{which leads to asymptotic convergence
for fading channels}.
%over the fading wireless channel and
%it is proven to converge asymptotically.
Achieving finite-time max-consensus
in the same wireless setting
is the topic of Section~\ref{sec:FTC}.
%The proposed switching protocol
%is proven to guarantee finite-time convergence.
Randomized simulations show the benefits of
this strategy \com{in comparison} to standard approaches.
Concluding remarks are stated
in Section~\ref{sec:concl}.
\subsection{Notation}
In the remainder
of this paper,
%$\mathbb{Z}$ is
%the integers set, 
%whilst
$\natn$ denotes
the set of nonnegative integers
and
$\mathbb{N}$ denotes
the set of positive integers.
The set of real numbers is denoted
by $\mathbb{R}$, while the set of
nonnegative, \com{respectively} positive, real numbers
by $\mathbb{R}_{\geq0}$, \com{respectively} $\mathbb{R}_{>0}$.
\com{
	A directed graph is a pair $(\mathcal{N},\mathcal{A})$,
	where $\mathcal{N}\subset\mathbb{N}$
	is the node set and $\mathcal{A}\subseteq\mathcal{N}\times\mathcal{N}$
	the arc set. $(i,j)\in\mathcal{A}$ is an arc
	from node $i$ to node $j$.
	A path from node $i_1$ to node $i_n$
	is a sequence of arcs
	$$
		(i_1,i_2), (i_2,i_3), \dots, (i_{n-1},i_n).
	$$
	The graph $(\mathcal{N},\mathcal{A})$
	is said to be \textit{strongly connected}
	if there is a path between all pairs of nodes.
}
%Complex numbers are grouped in
%set $\mathbb{C}$.
%The empty set is $\emptyset$.
%Let $\mathcal{N}\subset\mathbb{N}$ be a finite set,
%called \textit{node set}.
%Further, $\mathcal{A}$
%is called \textit{arcs set} and is defined
%as a subset of $\mathcal{N}\times\mathcal{N}$.
%The pair $(\mathcal{N},\mathcal{A})$
%denotes a graph;
%given two nodes $i,j\in\mathcal{N}$,
%the element $(i,j)\in\mathcal{A}$
%is an arc going from node $i$
%to node $j$.
%A graph is said to be undirected
%if $(i,j)\in\mathcal{A}$ implies
%$(j,i)\in\mathcal{A}$.
%Given a node $i$, 
%$N_i\subset\mathcal{N}$ is the set
%of its neighbors, i.e.
%$N_i:=
%\{ j\in\mathcal{N}\mid \{(j,i)\}\cap\mathcal{A}\not=\emptyset \}$.
%Given a graph $(\mathcal{N},\mathcal{A})$,
%a \textit{path} is a sequence of arcs of the form
%$\mathcal{o}_{i_1}^{i_n}=\{(i_1,i_2),(i_2,i_3)\dots(i_{n-1},i_n) \}$.
%An undirected graph
%is \textit{connected}
%if, given any pair $(i,j)\in\mathcal{A}$,
%there exists a path $\mathcal{o}_{i}^{j}$.
%Given sets $A$ and $B$ assumed to be finite,
%$A\setminus B$ is the set
%composed of all elements of
%$A$ that are not contained in $B$.
Given a finite set $A$,
its cardinality is denoted by $|A|$.
Given a set $A$, %where $\bar{a}$ and $\underline{a}$ are,
%respectively,
%its maximum and minimum elements,
$\mathcal{C}(A)$ denotes the convex hull
of $A$, i.e. 
$$\mathcal{C}(A):=\left\{ \sum_{i=1}^{|A|} \lambda_i x_i \mid {x_i} \in A, \lambda_i \geq 0,\ \sum_{i=1}^{|A|} \lambda_i=1 \right\}.$$
%$\mathcal{C}(A)=\{x\in\mathbb{R} \mid \underline{a} \leq x\leq\bar{a} \}$.

Given a vector $\mathbf{v}\in\mathbb{R}^n$, $n\in\mathbb{N}$,
its element in position
$i\in\{1,\dots,n\}$
is 
%$\mathbf{v}_i$
%or 
$[\mathbf{v}]_i$.
The transpose of $\mathbf{v}$ is $\mathbf{v}'$.
By $\mathbf{1}_n$, we denote the vector
composed of $n$ ones.
%
%Given a vector $\mathbf{v}\in\mathbb{C}^n$, $n\in\mathbb{N}$,
%its conjugate transpose is $\mathbf{v}^*$.
%Its norm is denoted by  $||\mathbf{v}||$, i.e.
%$$||\mathbf{v}||=\sqrt{\sum_{i=1}^n \mathbf{v}_i\mathbf{v}_i^*}\ .$$

Given $x\in\mathbb{R}$,
its absolute value is $|x|$,
$\lceil x\rceil$
denotes the least integer greater than or equal to $x$,
while $\lfloor x\rfloor$ is 
the greatest integer 
less than or equal to $x$. 
%Given a complex number $z\in\mathbb{C}$,
%$|z|$ denotes its modulus, while
%$\mathfrak{R}(z)$ its real part.
%
%
%If $p$ is a random variable, $\mathbb{E}\{p\}$ denotes
%its expected value,
%while $\mathrm{Var}\{p\}$ its variance.

The indicator function of a set $S\subseteq\mathbb{R}$,
denoted by
$I_S:\mathbb{R}\mapsto\{0,1\}$,
is defined by
$I_S(x)=1$ if $x\in S$
and $0$ otherwise.
%
%yields $1$
%if its argument is contained in 
%set $S\subseteq\mathbb{R}$,
%$0$ otherwise.
\section{System Description}
\label{sec:probDesc}
\subsection{Consensus in Multi-agent Systems}
\label{sec:conse}
We consider a multi-agent system with $n>1$ agents
communicating over the wireless channel
%%
%%Let the set $\mathcal{N}$ be a collection of $n\geq1$ agents
%%that communicate over a wireless network,
%We use $\mathcal{N}:=\{1\dots n\}$ to label the agents.
%All the communication links among agents are grouped in set $\mathcal{A}$,
%i.e., $\forall i,j\in\mathcal{N}$, $(i,j)\in\mathcal{A}$ if and only if a communication link from agent $i$ to agent $j$ exists.
%We assume that there are no self links, i.e.
%$\forall i\in\mathcal{N}$, $(i,i)\not\in\mathcal{A}$.
%%Therefore, the graph $(\mathcal{N}, \mathcal{A})$ models the multi-agent system, where
%%$\mathcal{N}$ and $\mathcal{A}$ will constitute, respectively, the set of nodes and the set of arcs.
%Therefore, the multi-agent system can be
and modeled by the directed
graph $(\mathcal{N}, \mathcal{A})$.
Given an agent $i\in\mathcal{N}$, 
$N_i\subset\mathcal{N}$ denotes the set of its neighbors, i.e.
$N_i:=\{j\in\mathcal{N} \mid (j,i)\in\mathcal{A} \}$.

%Each agent has a discrete-time dynamics and, in what follows, let $\kinN$ be the time index.
The multi-agent system seeks 
for an agreement over 
a variable of common interest.
Each agent $i\in\mathcal{N}$
has an initial estimation of this variable, which is referred to as its initial information state
\fa{{$x_{i_0}\in\mathrm{S}:=[S_{\mathrm{min}},S_{\mathrm{max}}]\subset\mathbb{R}_{\geq0}$}},
with $S$ being a compact set. 
In order to achieve \textit{consensus}, 
agents iteratively exchange information
with their neighbors and update their information states
according to a predefined \textit{consensus protocol}.
Let, $\forall i\in\mathcal{N}$, 
$\forall k\in\mathbb{N}_0$,
$\mathbf{x}_{N_i}(k)\in S^{|N_i|}$
be the set of information states 
of all agents in $N_i$ at 
iteration $\kinN$, i.e.,
\begin{equation}
	\label{eq:vector_neigh}
	\mathbf{x}_{N_i}(k):=
	[
		x_{j_1}(k),\dots,x_{j_{m_i}}(k)
	]',
\end{equation}
where $j_1,\dots,j_{m_i}\in N_i$ and $m_i=|N_i|$.
Widely considered discrete-time consensus protocols are of the form
\begin{equation}
	\label{eq:generalConsProt}
	x_i(k+1)=f_i(x_i(k),\mathbf{x}_{N_i}(k)),
\end{equation}
where $f_i:\mathrm{S}^{|N_i|+1}\rightarrow\mathrm{S}$ and, $\forall i\in\mathcal{N}$, $x_i(0)=x_{i_0}$.
The information states at iteration $\kinN$ are 
collected in the vector $\mathbf{x}(k)$, 
i.e., $\forall i\in\mathcal{N}$,
$[\mathbf{x}(k)]_i=x_i(k)$.
The system achieves consensus 
if $\exists x^*\in\mathrm{S}$ such that
\begin{equation}
	\label{eq:cons_achieved}
	\forall i\in\mathcal{N},\ 
	\lim\limits_{k\rightarrow\infty} x_i(k)=x^*.
\end{equation}
The system is said to achieve \textit{max-consensus} if
each information state converges to the largest initial information state of the multi-agent system, i.e.,
\begin{equation}
	\label{eq:max_cons_achieved}
	\forall i\in\mathcal{N},\ 
	\lim\limits_{k\rightarrow\infty} x_i(k)=x^*=
	\max_{i\in\mathcal{N}}({x}_{i_0}).
\end{equation}
If an agreement is achieved in a finite number of iterations,
then
we have \textit{finite-time max-consensus}. Formally, $\exists \bar{k}\in\natn$, such that
\begin{equation}
	\label{eq:def_finit_time_mC}
	\forall k>\bar{k},~\forall i\in\mathcal{N},\ x_i(k)=x^*=
	\max_{i\in\mathcal{N}}({x}_{i_0}).
\end{equation}
Any agent, say $i\in\mathcal{N}$, whose information state at 
iteration $\kinN$ is $x_i(k)=x^*=\max_{i\in\mathcal{N}}({x}_{i_0})$ is 
referred to as a \textbf{maximal agent} at iteration $k$.
%With regards to the general protocol (\ref{eq:generalConsProt}),
%max-consensus as defined in (\ref{eq:max_cons_achieved}), 
%can be reached only if,
%intuitively,
%\begin{equation}
%	\label{eq:necessaryForMax}
%	\forall \kinN,\ 
%	\max(\mathbf{x}(k))=\max(\mathbf{x}(0)).
%\end{equation}

\subsection{Max-Consensus: Benchmarking Protocol}
\label{sec:standard_approach}
In a \com{standard} max-consensus protocol,
each agent $i\in\mathcal{N}$ applies,
at every iteration $\kinN$,
the following \com{update}:
\begin{equation}
\label{eq:traditionalMaxCons}
	x_i(k+1)=\max_{ j\in N_i\cup \{i\} }
		(x_j(k)).
\end{equation}
Therefore, 
at iteration $\kinN$, each agent 
$i\in\mathcal{N}$ %is provided
%with the exact knowledge of neighbors' information states.
%By protocol (\ref{eq:traditionalMaxCons}),
%agent $i$ 
updates its information state 
by setting it to the largest 
value in the set 
$\{ x_i(k),x_{j_1}(k),\dots,x_{j_{m_i}}(k) \}$.
Under the assumption of a time-invariant and connected network topology,
\cite{nejad2009max} shows that max-consensus is achieved in
at most
$l\in\mathbb{N}$ steps, where
\begin{equation}
	\label{eq:standard_ftc}
	l=\max_{i,j=1,\dots,n}\{
		|i,j|_{l,\min}
	\},
\end{equation}
where $|i,j|_{l,\min}$ is the length of the shortest path connecting nodes $i$ and $j$.
Such a consensus protocol
\com{can be implemented if each agent has access to its neighbors' information states}.
%each agent requires the exact knowledge of neighbors' information states.
Under a communication point of view, this can be achieved by employing 
orthogonal channel access methods
together with error coding and re-transmission mechanisms
(for combatting noise), see \cite[Chapter 4]{Tse2012}.
%Accordingly, in the context of max-consensus problems over wireless channels,
For benchmarking purposes,
we \com{refer to the}
the joint usage of protocol (\ref{eq:traditionalMaxCons})
and orthogonal channel access methods
\com{as the \textit{standard approach for max-consensus}}.

In a wireless communication setting, 
orthogonal access to the communication channel can be ensured,
but may cause unacceptable costs in terms of 
higher overhead.
%
%although this requirement can be ensured,
%it carries some costs.
In fact, if the objective is 
to determine the maximum value in a given set,
it is in general 
\com{not necessary} to provide each agent with
the complete knowledge of each element in the set. % that set %the information states of neighbors 
%is a suboptimal strategy.
In fact,
by the \textit{data processing inequality} \cite{cover2012elements}, 
the amount of information contained in the set $\{ x_i(k),x_{j_1}(k),\dots,x_{j_{m_1}}(k) \}$
is in general 
larger than the amount of information carried by $\max_{ j\in N_i\cup \{i\} }(x_j(k))$.
Adopting orthogonal transmission 
(thus avoiding interference)
is therefore not necessary,
which motivates a more efficient 
method to achieve the same goal
by exploiting the interference.
% Navneet
%\import{./}{problem_communication.tex}
\subsection{{Wireless Multiple Access Channel} (WMAC)}
\label{sec:wirelessMAC}
The WMAC model describes
the communication
between multiple transmitters and a 
receiver over the fading wireless 
channel, see, e.g., \cite{ahlswede1973multi,Giridhar2006}.
All transmitters access the same channel
simultaneously.
The fading effect attenuates by a random
(and unknown)
coefficient all transmitted signals.
The receiver obtains
a superposition (sum) of such attenuated
signals.
\begin{definition}[WMAC]
	\label{def:MAC}
	Let $\mathcal{T} \subset \mathcal{N}$ 
	be a subset of agents transmitting
	to a designated agent 
	$i \in \mathcal{N}$.
	Each agent $j\in\mathcal{T}$ 
	at iteration $k\in\mathbb{N}$
	transmits a wireless signal
	$\omega_j(k) \in \mathbb{R}$.
	The signal obtained by the receiver is
	$z_i(k)\in\mathbb{R}$, computed as
	\begin{equation}
	\label{eq:WMAC}
	{
		z_i(k)
		:= \sum_{j\in\mathcal{T}} \xi_{ij}(k)\, \omega_j(k),
	}
	\end{equation}
	where, 
	$\forall \kinN$,
	$\forall j\in\mathcal{T}$,
	$\xi_{ij}(k)\in \mathbb{R}_{> 0}$ is the 
	real fading coefficient that captures the fading effect
	between the transmitter $j$ 
	and the receiver $i$.
\end{definition}
%Real fading coefficients $\xi_{ij}(k) \in \mathbb{R}_{> 0}$
%are modeled as independent and identically
%distributed 
%random numbers. They are drawn out of
%a Rayleigh or Rician distribution:
%the Rayleigh distribution is widely used to model non-line-of-sight scenarios,
%while the Rician distribution is assumed if a line-of-sight component is present. More details
%can be found in \cite[Ch 2]{Tse2012}.
%The real fading coefficients $\xi_{ij}(k)\in\mathbb{R}_{>0}$
%are modelled as independent and identically distributed random
%numbers.
%They are drawn out of a Rayleigh of Rician distribution\footnote{
%	The
%	Rayleigh distribution is widely used to model non-line-of-sight
%	scenarios, while the Rician distribution is assumed if a strong
%	line-of-sight component is present.
%}. More details can be found in
%\cite[Ch. 2.4]{Tse2012}.
%%
%In practice,
%fading coefficients
%and transmitted signals are
%complex valued. In order to obtain 
%(\ref{eq:WMAC}), we employ the communication system 
%designed by \cite{bjelakovic2019distributed}.
%In this contribution, $\forall i,j\in\mathcal{N}$,
%the real valued coefficient $\xi_{ij}$
%is taken as the square norm of the complex fading 
%channel between $j$ and $i$. More details
%can be found in \cite{bjelakovic2019distributed}.
%\newline
Under a communication theoretical point of view,
the wireless model presented in Definition~\ref{def:MAC}
is based on some assumptions.
\begin{assumption}[Wireless Channel]
	\label{assumption:channel}$ $
	\begin{itemize}
		\item[(A1)] The fading channel
		coefficients are assumed to be
		\emph{positive real numbers}.
		This is possible by employing a communication
		system as in \cite{goldenbaum2013robust}, 
		\cite{Kortke2014}, and \cite{bjelakovic2019distributed}.	
		\item[(A2)] The fading channel coefficients are assumed to be identically
		distributed and independent across different
		iterations and across different transmitter-receiver pairs.
		This is a valid assumption
		when the channel is \emph{fast-fading},
		see, e.g., \cite[Ch 2.3]{Tse2012}.
		As in \cite[Ch 2.4]{Tse2012},
		channel coefficients are drawn out of a Rayleigh or Rician distribution\footnote{
			The
			Rayleigh distribution is widely used to model non-line-of-sight
			scenarios, while the Rician distribution is assumed if a strong
			line-of-sight component is present.
		}.			
		\item[(A3)] The receiver is assumed
		to have no additive noise. Indeed, in a high-SNR (Signal
		to Noise Ratio) regime, 
		receiver noise can be neglected.
		\item[(A4)] Transmission and reception of
		wireless signals take place simultaneously across the network
		at every iteration $k\in\mathbb{N}$.
		This is possible by employing \emph{full-duplex transceivers}, 
		see, e.g., \cite{choi2010achieving}.
	\end{itemize}
%	%\begin{itemize}
%	%\item \textbf{Flat-fading and Real}
%	(A1)~The channel is fast-fading.
%	%\item \textbf{Noiseless} 
%	(A2)~The receiver is assumed to have no additive noise.
%	%\item \textbf{Independent and Identically distributed}
%	(A3)~Channel realizations are assumed to 
%	be identically distributed and
%	independent across different time instances
%	and across different transmitter-receiver pairs.
%	%\end{itemize}
%	(A4)~Transmission and reception
%	of wireless signals are simultaneous
%	across the network.
\end{assumption}
\subsection{Communication System}
\label{sec:commsys}
%The following assumption facilitates the construction of 
%WMACs with efficient usage of communication resources.
%
%\begin{assumption}[Full-Duplex and Synchronized Transceivers]
%\label{assumption:FD}
%We assume that every agent is equipped with a full-duplex transceiver system,
%capable of facilitating synchronized transmission and 
%reception of wireless signals. 
%\end{assumption}
%
%For our scheme, employing Half-Duplex transceivers,\footnote{Half-duplex transceivers can either transmit or receive at any time instant.}
%will lead to the communication resources (channel-uses) 
%required to share information among a set of fully-connected agents
%grow at least linear with increasing number of agents.
%%where a set of $N$ fully-connected agents will require at least $N$
%%channel uses (communication resources) to share information among themselves,
%Instead, employing Full-Duplex transceivers 
%will only require single channel-use to share information,
%independent of the number of participating agents.
%%since all $N$ agents can share information in a single channel use.
The goal of our communication system
is to provide each agent $i\in\mathcal{N}$
at every iteration $k\in\mathbb{N}_0$
with a signal $u_i(k)$
such that
\begin{equation}\label{eq:desiredComSys}
	u_i(k)\in\mathcal{C}
	\left(\{
	x_j(k)
	\}_{j\in N_i}\right),
\end{equation}
where $\mathcal{C}$ denotes the convex-hull and
$N_i$ is the set that includes
all agents transmitting to agent~$i$
(namely, its neighbors)\footnote{
	We assume $N_i$ to be non-empty.
}.
This is possible by designing
a communication system that processes
the signal both at each transmitter
and at the receiver.
%
%All transmitting agents $j\in\mathcal{T}$ are connected to receiver $i$
%and aim at communicating their information states.
%The commnications system's goal is to obtain (at the receiver)
%a value in the convex hull of set $\{x_j\}_{j\in\mathcal{T}}$.
%In the following, we present transmitter and receiver side processing
%of information that achieves system's goal in a WMAC.
% We assume that all transmitting agents $j \in \mathcal{T}$ are connected to
% the receiver $i$, and the goal of the communication system is to
% obtain, at the receiver $i$, a value in the convex hull
% of information states of the agents in set $\mathcal{T}$.
%i.e.~a value in the set defined as 
%$\mathcal{C} := \Big\{ \sum_{j \in \mathcal{T}} \lambda_j x_j : \forall j \in \mathcal{T},\ \lambda_j \geq 0,\ \sum_{j \in \mathcal{T}} \lambda_j = 1 \Big\} $.

\subsubsection{Transmitter-side processing}
% Each agent $j\in\mathcal{T}$ broadcasts its
% information state $(\forall k \in \mathbb{N}),\ x_j(k) \in \mathrm{S}$.
% to all of its neighbors $l \in N_j$.
All transmitters have the same power constraints
which restrict the amplitude of transmitted signals 
to a finite range, i.e., $\mathrm{P} := [P_{\min}, P_{\max}] \subset \mathbb{R}_{\geq0}$.
%In order to ensure transmitter-side power constraints,
Before transmission, the information state $x_j$ of an agent $j$
is transformed using an affine function
$\Phi: \mathrm{S} \to \mathrm{P}$
such that
\begin{equation}
\label{eq:scaling_function}
\Phi(x) = \alpha x + \beta,
\end{equation}
where 
$$
	\alpha = \frac{P_{\max} - P_{\min}}{S_{\max} - S_{\min}} \in \mathbb{R}_{>0}
$$ 
and 
$$
	\beta = P_{\min} - \alpha S_{\min}.
$$
The signal 
transmitted by agent $j$ at iteration
$k$ is
\begin{equation*}
	\mu_j(k) = \Phi(x_j(k)).
\end{equation*}
Also, as it will become 
clear in the receiver-side processing section,
in order for the receiver to normalize the fading coefficients,
a dummy signal $\mu'_j(k)$ %$(\forall k \in \mathbb{N}),\ \{x'_j(k) = 1\}_{j\in\mathcal{T}}$
is transmitted (orthogonal to $\mu_j(k)$).
Signal $\mu'_j(k)$ is given by, 
$\forall j \in N_i$, $\forall \kinN$,
\begin{equation*}
	\mu'_j(k) = \Phi(1) = \alpha+\beta.
\end{equation*}

\subsubsection{Receiver-side processing}
By the WMAC model (see Definition~\ref{def:MAC}),
each receiver $i\in\mathcal{N}$ obtains
two real-valued orthogonal signals at every iteration $k\in\mathbb{N}$, 
which are
\begin{equation}
\label{eq:ortho_recv}
\begin{aligned}
	r_i(k) &:= \sum_{j \in N_i} \xi_{ij}(k)\mu_j(k)
	= \sum_{j \in N_i} \xi_{ij}(k)(\alpha x_j(k) + \beta), \\ 
	r'_i(k) &:= \sum_{j \in N_i} \xi_{ij}(k)\mu'_j(k)
	= (\alpha + \beta)\sum_{j \in N_i} \xi_{ij}(k)   .
\end{aligned}
\end{equation}
At the receiver, 
we apply a de-scaling transformation
$\Psi:\mathbb{R}^2 \to \mathbb{R}$ 
that is
\begin{equation}
\label{eq:processed_num}
	\Psi(r_i, r'_i) :=  \frac{1}{\alpha} (r_i - \frac{\beta}{\alpha+\beta} r'_i)=\sum_{j \in N_i} \xi_{ij}(k)x_j(k),
\end{equation}
where $\alpha$ and $\beta$ are 
the values introduced in (\ref{eq:scaling_function}).
% The de-scaled signal $\Psi(r_i(k), r'_i(k))$ and signal $r'_i(k)$ 
% are divided to obtain the (normalized) desired value
% that is used by the controller as described in Section \ref{sec:asyCons}.
With this information at hand,
each agent $i\in\mathcal{N}$ can obtain
signal~(\ref{eq:desiredComSys}) at every iteration
$k\in\mathbb{N}_0$ by computing
\begin{equation}\label{eq:u_i_calc}
	u_i(k)=
	\cfrac{(\alpha+\beta)\Psi(r_i, r'_i)}{r'_i}.
\end{equation}
By inserting (\ref{eq:processed_num})
and (\ref{eq:ortho_recv}) into (\ref{eq:u_i_calc}),
one obtains
\begin{equation}
	u_i(k)=\sum_{j\in N_i} h_{ij}(k) x_j(k),
\end{equation}
where, $\forall i,j\in\mathcal{N}$,
$\forall k\in\mathbb{N}_0$,
$h_{ij}(k)\in\mathbb{R}_{}>0$ is referred to as the \textit{normalized
channel coefficient} corresponding to the transmission
from $j$ to $i$ at iteration $k$ and 
can be formally expressed as
\begin{equation}
	\label{Eq:normalizedFadChCoeff}
	h_{ij}(k) := \frac{\xi_{ij}(k)}{\sum_{q\in N_i} \xi_{iq}(k)}\ \in (0, 1].
\end{equation}
Note that each normalized channel coefficient
is unknown. However,
they sum up to $1$, i.e.,
$\forall i\in\mathcal{N}$, 
$\forall k\in\mathbb{N}_0$,
$$
	\sum_{j\in N_i}h_{ij}(k)=1.
$$
This proves that the requirement
(\ref{eq:desiredComSys})
is satisfied.
\section{Asymptotic Max-Consensus Protocol}
\label{sec:asyCons}
\subsection{Max-Consensus Protocol Design}
\com{In the following, we describe a max-consensus
protocol for the multi-agent system from Section~\ref{sec:conse}
and the communication system from Section~\ref{sec:commsys}.
It exploits interference and is }
based on two underlying ideas, 
which are summarized in Observation \ref{obs:not_need_max} and Proposition \ref{prop:local_eval}.
\begin{observation}
	\label{obs:not_need_max}
	If the goal is to achieve max-consensus, 
	any non-maximal agent at iteration 
	$\kinN$ 
	does not need to communicate 
	its own information state 
	at iteration $\kinN$.
\end{observation}
%
%As in (\ref{eq:max_cons_achieved}), 
%the agreement value unquestionably depends only on $\max(\mathbf{x}(0))$.
%Intuitively, in the context of a max-consensus problem, 
%the information carried at iteration $\kinN$ 
%by non-maximal agents at iteration $k$ is not needed.
%However, in general, agents do not have the overall knowledge of the system. 
\com{However, agents in general do
not know whether they are maximal at a given iteration $k$}.
Agents only have a local estimation of this. 
Let us assume that the result of 
this local evaluation for agent $i\in\mathcal{N}$
at iteration $\kinN$
is stored in a binary variable
$\tilde{y}_i(k)\in\{0,1\}$.
In the case $\tilde{y}_i(k)=1$, 
agent $i\in\mathcal{N}$ 
is said to be a \textbf{maximal-candidate} at iteration $k$.
If (and only if) 
agent $i$ is a maximal-candidate, 
it will be allowed to broadcast at the next iteration.
%according to the evidence presented in Observation \ref{obs:not_need_max}.
This will be expressed by an \textbf{authorization variable} 
$y_i:\natn\rightarrow\{0,1\}$, where $y_i(k)=\tilde{y}_i(k-1)$
\com{
	and $y_i(0)=1$
}.

%The idea can be formalized as follows. 
Now, let,
\begin{multline}
	\label{eq:neighAuth}
	\forall i\in\mathcal{N},~\forall \kinN,
	\\
	N_i^m(k):=\{ j\in N_i \mid y_j(k)=1  \}\subseteq {N}_i
\end{multline}
be the set of neighbors authorized to broadcast at iteration $k$.
If only authorized 
agents broadcast %(see (\ref{eq:neighAuth}))
and the communication system of Section \ref{sec:commsys}
is employed, %(see (\ref{eq:u_avgCons})),
each agent $i\in\mathcal{N}$ will receive, at $\kinN$,
%\begin{equation}
%	\label{eq:z_i_allow}
%	z_i(k)
%	=\sum_{j\in N_i} h_{ij}(k)y_j(k)x_j(k)
%	=\sum_{j\in N_i^m(k)} h_{ij}(k)x_j(k)
%\end{equation}
%and
%\begin{equation}
%	\label{eq:z_i_p_allow}
%	z_i'(k)
%	=\sum_{j\in N_i} h_{ij}(k)y_j(k)
%	=\sum_{j\in N_i^m(k)} h_{ij}(k).	
%\end{equation}
%These two pieces of information will allow agent $i\in\mathcal{N}$ to locally evaluate if it is a maximal-candidate at time $k$. 
%In fact, let $\forall i\in\mathcal{N},\ \forall \kinN$, the real-valued nonnegative signal $u_i(k)$ be defined as
\begin{equation}
	\label{eq:u_i}
	u_i(k) =
%	\begin{cases}
%		\hat{f}_i(\mathbf{x}_{N_i^m}(k))
%		=
		{\sum\limits_{j\in N_i^m(k)} h_{ij}(k)x_j(k)}.
%		&\text{if }N_i^m(k)\not=\emptyset
%		\\
%		0 &\text{else.}
%	\end{cases}
%	.
\end{equation}
The local evaluation
that establishes which agents
are authorized to broadcast
is based on the following \com{proposition}.
\begin{proposition}
	\label{prop:local_eval}
	Given a set of agents $\mathcal{N}$, 
	a non-empty subset $\mathcal{M}\subseteq\mathcal{N}$, 
	and a set of real-valued parameters $\mathcal{H}=\{ h_j\in(0,1] \mid j\in\mathcal{M} \}$
	with $\sum_{j\in\mathcal{M}}h_j=1$,
	the following holds
	$\forall \kinN$, $\forall i\in\mathcal{N}$,
	\begin{equation}
	\label{eq:necCond}
	x_i(k)<{\sum_{j\in\mathcal{M}} h_j x_j(k)}
	\implies
	x_i(k)<\max_{j\in\mathcal{N}}(x_j(k)).
	\end{equation}
	\begin{proof}
		\com{By definition of a convex hull},
%		It is immediate to show that
%		$\forall\mathcal{M}\subseteq\mathcal{N}$,
		%		$\forall\mathcal{H}$,
		${\sum_{j\in\mathcal{M}} h_j x_j(k)}\in\mathcal{C}(\{x_j(k)\mid j\in\mathcal{M} \}).$
%		By definition of convex hull, 
		\com{Moreover,}
		$\forall p\in\mathcal{C}(\{x_j(k)\mid j\in\mathcal{M} \}):\ p\leq \max_{j\in\mathcal{M}}(x_j(k))$. 
%		This together with the cond<ition on
%		the left-hand side of (\ref{eq:necCond}), yields
		\com{Hence},
		\begin{equation*}
%		x_i(k)<
		{\sum_{j\in\mathcal{M}} h_j x_j(k)}
		\leq \max_{j\in\mathcal{M}}(x_j(k)).
		\end{equation*}
%		which 
%		implies $x_i(k)<\max_{j\in\mathcal{M}}(x_j(k))$.
		Since $\mathcal{M}\subseteq\mathcal{N}$, 
		$$
			x_i(k)<\max_{j\in\mathcal{M}}(x_j(k))\leq\max_{j\in\mathcal{N}}(x_j(k))
			.
		$$
%		from which (\ref{eq:necCond}) immediately follows.
		\com{This implies (\ref{eq:necCond})}.
	\end{proof}
\end{proposition}

By (\ref{eq:necCond}), for $\mathcal{M}=N_i^m(k)$ and $\mathcal{H}=\{ h_{ij}(k) \mid j\in N_i^m(k) \}$,
the \com{implication}
\begin{equation}
	\label{eq:applyingCond}
	x_i(k)<u_i(k)\implies x_i(k)<\max_{j\in\mathcal{N}}({x}_j(k))
\end{equation}
immediately follows. Therefore, $y_i$ can be
updated as
\begin{multline}
	\label{eq:updateY} 
	\forall i\in\mathcal{N},~\forall \kinN,\\
	y_i(k+1)=\tilde{y}(k)=
	I_{\mathbb{R}\geq0}(x_i(k)-u_i(k)).
\end{multline}
In the light of these observations, 
and given that the signal $u_i(k)$ is computed by harnessing the interference of the channel,
each agent $i\in\mathcal{N}$ can apply
the following max-consensus protocol:
\begin{equation}
	\label{eq:consProtAsympt}
	\forall \kinN,\ 
	\begin{cases}
		x_i(k+1)=\max(x_i(k),u_i(k))\\
		y_i(k+1)=I_{\mathbb{R}_{\geq0}}(x_i(k)-u_i(k))
	\end{cases}
	,
\end{equation}
where $y_i(0)=1$ and $x_i(0)=x_{i_0}\in \mathrm{S}$, 
and $u_i(k)$ is obtained %$\forall \kinN$ 
from (\ref{eq:u_i}).
\com{Note that} $u_i(k)$ \com{is determined by} 
$x_j(k)$ and $y_j(k)$, $j\in N_i$. 
%By this, 
(\ref{eq:u_i})-(\ref{eq:consProtAsympt}) can
then be rewritten in vector-form as
\begin{equation}
	\label{eq:consMatForm}
	\mathbf{w}(k+1)
	=
	g(
	\mathbf{w}(k)
	),
\end{equation}
where
\begin{equation}
	\label{w}
	\mathbf{w}(k)
	=
	\left[
	\begin{matrix}
		\mathbf{x}(k)\\
		\mathbf{y}(k)
	\end{matrix}
	\right],
\end{equation}
and, $\forall i\in\mathcal{N},
~[\mathbf{x}(k)]_i=x_i(k),
~[\mathbf{y}(k)]_i=y_i(k)$,
and
$g:\mathbb{R}^{n}\times\{0,1\}^n\rightarrow\mathbb{R}^n\times\{0,1\}^n$
is the nonlinear function reflecting~(\ref{eq:consProtAsympt})
and (\ref{eq:u_i}).

\subsection{Asymptotic Convergence of the System}
A multi-agent system with a 
\com{strongly}
connected network topology $(\mathcal{N},\mathcal{A})$ is given.
The system uses
the consensus protocol (\ref{eq:consMatForm}),
\com{i.e.,} each agent iterates
(\ref{eq:u_i}),(\ref{eq:consProtAsympt}) synchronously.
In the following, we prove
%the 
asymptotic 
%max-consensus 
convergence %of the system.
by using Lyapunov theory 
(cf. \cite[p. 87]{aastrom2013computer} and \cite[p. 22]{lalo2014advanced}).
Initially, we show that all information states
%evolve according 
\com{are}
%to a 
non-decreasing bounded sequences.
\begin{proposition}
	\label{prop:nondec}
	Given a multi-agent system with network topology $(\mathcal{N},\mathcal{A})$
	\fa{and} consensus protocol (\ref{eq:consMatForm}),
	$\forall \mathbf{x}(0)\in\mathrm{S}^n$, $\forall i\in\mathcal{N}$, $\forall \kinN$, 
	\begin{equation}
		\label{eq:nonDec}
		x_i(k)\leq x_i(k+1)\leq \max_{j\in\mathcal{N}}({x}_j(0)).
	\end{equation}
	\begin{proof}
		The first inequality immediately follows from (\ref{eq:consProtAsympt}). 
		The second inequality follows
		\com{
		from the fact that, according to~(\ref{eq:u_i}),
		$$
			u_i(k)\in\mathcal{C}\left(
				\{
					x_j(k)
					\mid
					N_i^m(k)
				\}
			\right)
		$$
		if $N_i^m(k)\not=\emptyset$, zero else.
		Hence
		$$
			u_i(k)
			\leq
			\max_{j\in N_i^m(k)}\left(x_j(k)\right)
			\leq
			\max_{j\in \mathcal{N}}\left(x_j(k)\right).
		$$
		Therefore,
		$\forall i\in\mathcal{N}$,
		$$
			x_i(k+1)
			\leq
			\max_{j\in N_i^m(k)}\left(x_j(k)\right)
			\leq
			\max_{j\in \mathcal{N}}\left(x_j(k)\right).
		$$
		Moreover,
		$$
			\max_{j\in \mathcal{N}}\left(x_j(k+1)\right)
			\leq
			\max_{j\in \mathcal{N}}\left(x_j(k)\right),
		$$
		}
%		Indeed, $\forall i\in\mathcal{N}$, $\forall \kinN$, 
%		$x_i(k+1)=\max(x_i(k),u_i(k))\geq x_i(k).$
%		As for the second inequality, as already shown in the proof of Proposition \ref{prop:local_eval}, 
%		$\forall p\in\mathcal{C}(\{x_j(k)\mid j\in\mathcal{N} \})$, $p\leq\max_{j\in \mathcal{N}}(x_j(k))$.
%		By (\ref{eq:u_i}), $u_i(k)\in\{0\}\cup\mathcal{C}(\{x_j(k)\mid j\in\mathcal{N} \})$, 
%		that yields, $\forall i\in\mathcal{N},\ \forall \kinN,\ u_i(k)\leq\max_{j\in \mathcal{N}}(x_j(k))$.
%		By this, 
%		$\forall i\in\mathcal{N}$,
%		$\forall \kinN$,
%		$\max(x_i(k),u_i(k))\leq\max(\mathbf{x}(k)),$
		thus yielding the second inequality.
	\end{proof}
\end{proposition}
The following propositions
\com{establish a unique equilibrium point}.
\begin{proposition}
	\label{prop:eqSys}
%	A multi-agent system $(\mathcal{N},\mathcal{A})$
%	with a connected network topology
%	employs
%	protocol (\ref{eq:consMatForm}).
%	An equilibrium point for the system is
	\com{
	\begin{equation}	
		\label{eq:equilib}
		\mathbf{w}^*=
		\left[
			{\mathbf{x}^*}',
			\mathbf{1}_n'
		\right]',
	\end{equation}
	with $\mathbf{x}^*=x^*\mathbf{1}_n$ and $x^*=\max_{j\in\mathcal{N}}({x}_j(0))$
	is an equilibrium point
	for the multi-agent system with network
	topology $(\mathcal{N},\mathcal{A})$
	and consensus protocol (\ref{eq:consMatForm}).
	}
	\begin{proof}
		\com{
			Assume		
			$\mathbf{w}(k)=\mathbf{w}^*$.
			This implies,
			$\forall i\in\mathcal{N}$,
			$x_i(k)=x^*$
			and 
			$y_i(k)=1$.
			Therefore
			$
			N_i^m(k)=N_i$,
			hence
			\begin{align}
				u_i(k) &= \sum_{j\in N_i} h_{ij}(k)x_j(k)\nonumber\\
				&=\left(\sum_{j\in N_i} h_{ij}(k)\right)x^*
				=x^*,
			\end{align}
%				\begin{cases}
%					x^*\qquad\text{if }N_i\not=\emptyset\\
%					0\qquad\text{else}
%				\end{cases}.
			as $N_i\not=\emptyset$. 
			Then, according to (\ref{eq:consProtAsympt}),
			\begin{align}
				x_i(k+1)&=\max(x^*, u_i(k))=x^*\\
				\intertext{and}
				y_i(k+1)&=I_{\mathbb{R}_{\geq 0}}\left(
					x^* - u_i(k)
				\right)=1,
			\end{align}
			and, therefore, 
			$\mathbf{w}(k+1)=\mathbf{w}^*$.
		}
%		Evidently, $\mathbf{w}^*=[{\mathbf{x}^*}', {\mathbf{y}^*}']'$ is an equilibrium point for (\ref{eq:consProtAsympt}) 
%		if and only if, $\forall i\in\mathcal{N}$, 
%		\begin{align}
%			\label{eq:first_con_eq}
%			[\mathbf{x}^*]_i&=\max([\mathbf{x}^*]_i,[\mathbf{u}^*]_i)\\
%			\label{eq:second_con_eq}
%			[\mathbf{y}^*]_i&=I_{\mathbb{R}_{\geq0}}([\mathbf{x}^*]_i-[\mathbf{u}^*]_i) ,
%		\end{align}
%		where the vector $\mathbf{u}^*\in\mathbb{R}_{\geq0}^n$ 
%		is defined through each element 
%		$[\mathbf{u}^*]_i\in\mathbb{R}_{\geq0}$, 
%		$i\in\mathcal{N}$,  
%		so that
%		\begin{equation}
%			[\mathbf{u}^*]_i=
%%			\begin{cases}
%				{\sum_{j\in N_i^m} h_{ij}[\mathbf{x}^*]_j}%{\sum_{j\in N_i^m} h_{ij}}
%%				&\text{if }N_i^m\not=\emptyset
%%				\\
%%				0 &\text{else,}
%%			\end{cases}
%			,
%		\end{equation}
%		for a collection of $h_{ij}\in(0,1]$,
%		such that $\forall i\in\mathcal{N}$,
%		$\sum_{j\in N_i}h_{ij}=1$.
%		For ${\mathbf{y}^*}=\mathbf{1}_n$, $N_i^m=N_i$, which is therefore non-empty, since $(\mathcal{N},\mathcal{A})$ connected.
%		Then, $\forall i\in\mathcal{N}$,
%		\begin{equation}
%			[\mathbf{u}^*]_i={\sum_{j\in N_i} h_{ij}[\mathbf{x}^*]_j}.
%		\end{equation}
%		For $\mathbf{x}^*=x^*\mathbf{1}_n$ (i.e., $\forall i\in\mathcal{N},\ [\mathbf{x}^*]_i=x^*$), any possible collection of channel coefficients $h_{ij}\in(0,1]$ 
%		yields $\forall i\in\mathcal{N},\ [\mathbf{u}^*]_i=x^*$.
%		Consequently, (\ref{eq:first_con_eq}) and (\ref{eq:second_con_eq}) are satisfied.
	\end{proof}	
\end{proposition}
\begin{proposition}
	\label{cor:unicity_eq}
	\com{
		Consider a multi-agent system
		with a strongly connected network
		topology $(\mathcal{N},\mathcal{A})$
		using the consensus protocol (\ref{eq:consProtAsympt}).
		$$
			\mathbf{w}^*=
			\left[
			{x}^*\mathbf{1}_n',
			\mathbf{1}_n'
			\right]'
		$$
		is the unique equilibrium point.
	}
%	A multi-agent system with a 
%	connected network topology $(\mathcal{N},\mathcal{A})$ 
%	uses protocol 
%	(\ref{eq:consMatForm}). 
%	Then, $\forall \mathbf{x}(0)\in\mathrm{S}^n$, $\mathbf{w}^*$ is the unique equilibrium point of the system.
	\begin{proof}
		\com{
			The proof is by contradiction,
			i.e., we assume that there exists an equilibrium point
			$$
				\hat{\mathbf{w}}=
				[\hat{\mathbf{x}}',\hat{\mathbf{y}}']'
				\not=
				\mathbf{w}^*.
			$$			
			\paragraph{Case 1}
			$\hat{\mathbf{y}}\not=\mathbf{1}_n'$,
			i.e.,
			$$
				\exists i\in\mathcal{N},~\text{s.t.}~\hat{\mathbf{y}}_i=0.
			$$
			Hence, because of (\ref{eq:consProtAsympt}),
			$$
				x_i(k)<u_i(k)
			$$
			and therefore $x_i(k+1)>x_i(k)$.
			Hence we have established that for any equilibrium
			point $\hat{\mathbf{w}}$
			the Boolean part needs
			to be 
			$$ 
				\hat{\mathbf{y}}=\mathbf{1}_n'.
			$$
			\paragraph{Case 2}
			$ 
			\hat{\mathbf{y}}=\mathbf{1}_n'.
			$ 
			but
			$
			\hat{\mathbf{x}}\not=x^*\mathbf{1}_n'
			$.
			Because of the first premise,
			$\forall i\in\mathcal{N}$,
			$$
				N_i^m(k)=N_i.
			$$	
			As
			$(\mathcal{N},\mathcal{A})$
			is strongly connected,
			$N_i\not=\emptyset$, $\forall i\in\mathcal{N}$.
			Furthermore,
			also because of
			$(\mathcal{N},\mathcal{A})$
			being strongly connected,
			there exists a minimal element
			$l\in\mathcal{N}$
			that has at least one non-minimal
			neighbor, i.e.,
			$$
				\hat{x}_l=\min_{j\in\mathcal{N}}\hat{x}_j
			$$
			and
			$$
				\hat{x}_l<\max_{j\in N_l}\hat{x}_j:=\hat{x}_p.
			$$
			As the channel coefficients are positive,
			(\ref{eq:u_i}) implies
			$$
				u_l(k)>x_l(k).
			$$
			This and (\ref{eq:consProtAsympt})
			imply
			$$
				x_l(k+1)>x_l(k).
			$$
			Hence, we have established that for any equilibrium
			point $\hat{\mathbf{w}}$, the real part
			has to be $x^*\mathbf{1}_n$.
		}
	\end{proof}
\end{proposition}
%The following result allows to apply the Lyapunov theory. 
\begin{lemma}
	\label{prop:increase2steps}
	\com{Consider a} multi-agent system with a
	\com{strongly} connected network topology $(\mathcal{N},\mathcal{A})$.
	If protocol (\ref{eq:consMatForm}) is employed,
	then, $\forall \mathbf{x}(0)\in\mathrm{S}^n$, $\forall \kinN$,
	\begin{equation}\label{eq:xkp2}
		\sum_{i\in\mathcal{N}}\left( x_i(k+2) - x_i(k) \right) = 0 \implies \mathbf{x}(k)=\mathbf{x}^*.
	\end{equation}
	\begin{proof}
		\com{From} Proposition \ref{prop:nondec}, 
		$\{x_i(k)\}_{\kinN}$ is a non-decreasing bounded sequence, composed of nonnegative entries.
		As a consequence, $\sum_{i\in\mathcal{N}}\left( x_i(k+2) - x_i(k) \right)=0$ if and only if
		\begin{equation}
			\label{eq:propIncr_x0_x2}
			\mathbf{x}(k)=\mathbf{x}(k+1)=\mathbf{x}(k+2).
		\end{equation}
		The latter, by (\ref{eq:consProtAsympt}), implies that,
		$\forall i\in\mathcal{N},\ x_i(k)\geq u_i(k)$ and $x_i(k+1)\geq u_i(k+1)$.
		By (\ref{eq:updateY}), this implies that
		\begin{equation}
			\label{eq:propIncr_yIncr}
			\mathbf{y}(k+1)=\mathbf{y}(k+2)=1.
		\end{equation}
		From (\ref{eq:propIncr_x0_x2}) and (\ref{eq:propIncr_yIncr}), it clearly follows that 
		\begin{equation}
			\mathbf{w}(k+1)
			=
			\begin{bmatrix}
				\mathbf{x}(k+1)\\
				\mathbf{y}(k+1)
			\end{bmatrix}
			=
			\begin{bmatrix}
				\mathbf{x}(k+2)\\
				\mathbf{y}(k+2)
			\end{bmatrix}
			=	
			\mathbf{w}(k+2)
		\end{equation} 
		is an equilibrium for the system. 
		According to Proposition \ref{cor:unicity_eq}, 
		\com{it is unique}, i.e.,
		\begin{equation}
			\mathbf{w}(k+1)
			=	
			\mathbf{w}(k+2)
			=	
			\mathbf{w}^*,
		\end{equation} 
%		which implies that
		\begin{equation}
			\mathbf{x}(k+1)
			=	
			\mathbf{x}(k+2)
			=	
			\mathbf{x}^*.
		\end{equation} 
		By (\ref{eq:propIncr_x0_x2}), $\mathbf{x}(k)=\mathbf{x}^*$; this concludes the proof.
	\end{proof}
\end{lemma}
\begin{corollary}
	\label{cor:increase2steps_strict}
	\com{Consider a} multi-agent system with a
	\com{strongly} connected network topology 
	$(\mathcal{N},\mathcal{A})$.
	If the protocol (\ref{eq:consMatForm}) is employed,
	then, $\forall \mathbf{x}(0)\in\mathrm{S}^n$, $\forall \kinN$,
	\begin{equation}\label{eq:increasing}
		\mathbf{x}(k)\not=\mathbf{x}^* \implies \sum_{i\in\mathcal{N}}\left( x_i(k+2) - x_i(k) \right) > 0.
	\end{equation}
	\begin{proof}
%		It follows immediately from Proposition \ref{prop:increase2steps} and Proposition \ref{prop:nondec}.
		\com{
			(\ref{eq:xkp2}) is equivalent to 
			$$
				\mathbf{x}(k)\not=\mathbf{x}^*
				\implies
				\sum_{i\in\mathcal{N}}
				(x_i(k+2)-x_i(k))\not=0.
			$$	
			Non-decreasingness of the sequence
			$\{x_i(k)\}_{k\in\mathbb{N}_0}$
			(Proposition~\ref{prop:nondec})
			then establishes (\ref{eq:increasing}).
	}
	\end{proof}
\end{corollary}

%
%With these results in hand, 
%it is then possible to formalize the asymptotic convergence of the system to max-consensus.
By \cite[p. 264]{vidyasagar2002nonlinear} and \cite[p. 43]{murray2017mathematical},
a Lyapunov-based analysis can be applied to the
discrete-time system like (\ref{eq:consMatForm}), 
as existence and uniqueness 
of an equilibrium point \com{have been established in}
Proposition \ref{prop:eqSys} and Proposition \ref{cor:unicity_eq}. 
\begin{theorem}
	\label{prop:Lyapunov}
	\com{Consider a} multi-agent system with a
	\com{strongly} connected network topology $(\mathcal{N},\mathcal{A})$. Agents employ
	the consensus protocol (\ref{eq:consMatForm}).
	For every possible initial state $\mathbf{x}(0)\in\mathrm{S}^n$, 
	the system achieves max-consensus asymptotically.
	\begin{proof}
%		First, we investigate a possible way to rewrite the consensus protocol (\ref{eq:consMatForm}).
		The component $y_i$ of (\ref{eq:consProtAsympt}) can be explicitly rewritten, $\forall \kinN$, as
		\begin{align}
			\label{eq:lyap_y_expl}
			y_i(k+1)
			=
			\begin{cases}
				1 &\text{if }x_i(k)\geq u_i(k)\\
				0 &\text{if }x_i(k)< u_i(k)
			\end{cases}.
		\end{align}
		By (\ref{eq:consProtAsympt}) and since $\{x_i(k)\}_{\kinN}$ is a non-decreasing sequence, (\ref{eq:lyap_y_expl}) can be 
		reformulated, $\forall \kinN$, as
		\begin{align}
			\label{eq:lyap_y_expl_2}
			y_i(k+1)
			=
			\begin{cases}
				1 &\text{if }x_i(k)= x_i(k+1)\\
				0 &\text{if }x_i(k)< x_i(k+1)
			\end{cases}.
		\end{align}
		\com{This,
		again because of non-decreasingness
		of $\{x_i(k)\}_{k\in\mathbb{N}_0}$,
		is equivalent to
		}
		\begin{align}
			\label{eq:lyap_y_impl}
			\forall \kinN,\ 
			y_i(k+1)
			=
			I_{\mathbb{R}_{\geq0}}
				(x_i(k)-x_i(k+1)).
		\end{align}
		\com{
		By introducing the new state vector
		\begin{equation}
			\label{eq:stateVector}
			\mathbf{v}(k):=
			\left[
				\begin{matrix}
					\mathbf{v}_1(k)\\
					\mathbf{v}_2(k)
				\end{matrix}
			\right]:=
			\left[
				\begin{matrix}
					\mathbf{x}(k)\\
					\mathbf{x}(k-1)
				\end{matrix}
			\right].
		\end{equation}
		we can rewrite (\ref{eq:consMatForm})
		as
		\begin{equation}
			\label{eq:lyap_mat_form_passcoord}
			\mathbf{v}(k+1)=\tilde{g}(\mathbf{v}(k)),
		\end{equation}
		where the function $\tilde{g}:S^{2n}\mapsto S^{2n}$
		can be explicitly expressed by
		\begin{align}
			\mathbf{v}_1(k+1)&=\max(\mathbf{v}_1(k), H(k)\mathbf{v}_1(k))\\
			\mathbf{v}_2(k+1)&=\mathbf{v}_1(k),
		\end{align}
		where $\max(\cdot)$ is understood element-wise
		and
		$$
			[H(k)]_{ij}:=
			\begin{cases}
				h_{ij}(k)\qquad\text{if }j\in N_i~\text{and}~ [\mathbf{v}_1(k)]_j=[\mathbf{v}_2(k)]_j\\
				0\qquad\qquad\text{else}
			\end{cases}.
		$$
		}%
%		The latter leads to rewrite the system dynamics (\ref{eq:consMatForm}) as
%		\begin{equation}
%			\label{eq:lyap_mat_form_passcoord}
%			\mathbf{v}(k+1)=\tilde{g}(\mathbf{v}(k)),
%		\end{equation}
%		with $\tilde{g}:\mathrm{S}^{2n}\rightarrow\mathrm{S}^{2n}$ and 
%		the state vector
%		\begin{equation}
%			\label{eq:stateVector}
%			\mathbf{v}(k):=
%			\left[
%				\begin{matrix}
%					\mathbf{x}(k)\\
%					\mathbf{x}(k-1)
%				\end{matrix}
%			\right].
%		\end{equation}
		Consequently, the unique
		equilibrium $\mathbf{w}^*$ of (\ref{eq:consMatForm}) corresponds to the following unique
		equilibrium $\mathbf{v}^*$ of 
		(\ref{eq:lyap_mat_form_passcoord}):
		\begin{equation}
			\label{eq:lyap_new_eq}
			\mathbf{v}^*=x^*\mathbf{1}_{2n}.
		\end{equation}
		A candidate Lyapunov function $V:\mathrm{S}^{2n}\rightarrow\mathbb{R}_{\geq0}$ is chosen
		as follows:
		\begin{align}
			\forall \kinN,\
			V(\mathbf{v}(k))
			&=\mathbf{1}_{2n}(\mathbf{v}^*-\mathbf{v}(k))
			\\			
			\label{eq:lyap_func}
			&=2nx^*-\sum_{i\in\mathcal{N}}(x_i(k)+x_i(k-1)).
		\end{align}
		The following properties hold:
		\begin{itemize}
			\item[a)] $V(\mathbf{v})$ is continuous in $\mathbb{R}_{\geq0}^{2n}$;
			\item[b)] $V(\mathbf{v}^*)=0$;
			\item[c)] $V(\mathbf{v})$ is positive on any trajectory $\mathbf{v}$ of the system, unless $\mathbf{v}=\mathbf{v}^*$. 		
				
			This follow directly from Proposition \ref{prop:nondec}.
%			, in fact, 
%			$\forall i\in\mathcal{N},\ x_i\leq x^*$, therefore $V(\mathbf{v})\geq0$.
%			However, $V(\mathbf{v})=0$ only if,
%			$\forall i\in\mathcal{N},\ x_i=x^*$, 
%			that, by (\ref{eq:def_finit_time_mC}), is when the system has achieved max-consensus.
			\item[d)] By Corollary \ref{cor:increase2steps_strict}, $\forall \mathbf{v}(k)\not=\mathbf{v}^*$
			(%which is clearly equivalent to $\forall
			\com{i.e.,} $\forall\mathbf{x}(k-1)\not=\mathbf{x}^*$),
			\begin{align*}
				%\label{eq:decreasing_lyap}
				\Delta V(\mathbf{v}(k)) 
				&= V(\mathbf{v}(k+1))-V(\mathbf{v}(k))\\
				&= -\sum_{i\in\mathcal{N}}
				\left(
					x_i(k+1)-x_i(k-1)				
				\right)<0.
			\end{align*}
		\end{itemize}
		\com{Hence, the} function $V(\mathbf{v})$ 
		\com{has all the properties required for a Lyapunov function}.
		By \cite[p. 22]{lalo2014advanced} and \cite[p. 88]{aastrom2013computer}, 
		the system therefore asymptotically converges to max-consensus.
	\end{proof}
\end{theorem}
%Given such a convergence result, we are now interested in characterizing the converging behavior of each single agent. 
%The following result will be of use in the next section.
In the following corollary, an immediate result coming
as a consequence from Theorem \ref{prop:Lyapunov}
is reviewed. It
will be used in the next section.
\begin{figure}[t]
	\centering
	\begin{subfigure}[t]{.48\columnwidth}
		\centering
		\resizebox{\textwidth}{!}{%
			\begin{tikzpicture}
	\begin{scope}[every node/.style={circle,thick,draw,fill=orange,text=black}]
	    \node[label={\small a}] (A) at (0,0) {3};
	    \node[label={[label distance=0cm]10:\small b}] (B) at (1.5,1) {4};
	    \node[label={[label distance=0cm]-10:\small c}] (C) at (1.5,-1) {3};
	    \node[label={\small d}] (D) at (3,0) {3};
	\end{scope}
	
	\begin{scope}[>={Stealth[black]},
	              every node/.style={fill=white,circle},
	              every edge/.style={draw=black,very thick}]
	    \path [<->] (A) edge (B);
	    \path [<->] (B) edge (C);
	    \path [<->] (A) edge (D);
	    \path [<->] (D) edge (C);
	    \path [<->] (A) edge (C);
	    \path [<->] (B) edge (D);
	\end{scope}
\end{tikzpicture}%
		}
		\caption{Network topology for Examples 1, 2, 4. 
			The node set is $\{a,b,c,d\}$ and the vector of initial information states is $\mathbf{x}(0)=[3,4,3,3]'$
			for Example~1 and 4, and
			$\mathbf{x}(0)=[3.1,4,3,3]'$
			for Example 2.
			}
		\label{fig:graph_as_conv}
	\end{subfigure}\hfill%
	\begin{subfigure}[t]{.48\columnwidth}
		\centering
		\resizebox{\textwidth}{!}{%
			\begin{tikzpicture}
	\begin{scope}[every node/.style={circle,thick,draw,fill=orange,text=black}]
	    \node[label={\small a}] (A) at (0,0) {3};
	    \node[label={[label distance=0cm]10:\small b}] (B) at (1.5,1) {4};
	    \node[label={[label distance=0cm]-10:\small c}] (C) at (1.5,-1) {3};
	    \node[label={\small d}] (D) at (3,0) {3};
	\end{scope}
	
	\begin{scope}[>={Stealth[black]},
	              every node/.style={fill=white,circle},
	              every edge/.style={draw=black,very thick}]
	    \path [<->] (A) edge (B);
	    \path [<->] (B) edge (C);
	    \path [<->] (A) edge (D);
%	    \path [<->] (D) edge (C);
	    \path [<->] (A) edge (C);
	    \path [<->] (B) edge (D);
	\end{scope}
\end{tikzpicture}%
		}
		\caption{Network topology for Example 3. The node set is $\{a,b,c,d\}$ and the vector of initial information states is $\mathbf{x}(0)=[3,4,3,3]'$.}
		\label{fig:graph_as_conv_topDiff}
	\end{subfigure}
	\caption{Network topologies considered in Examples 1-4.}
\end{figure}
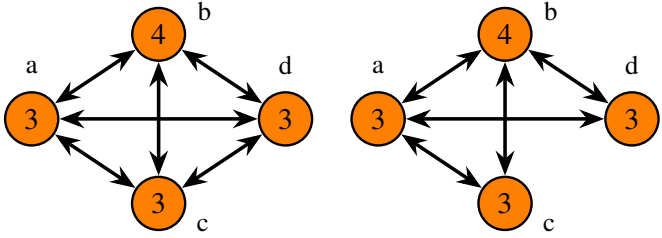
\begin{figure}[t]
	\centering
	\includegraphics[width=\columnwidth]{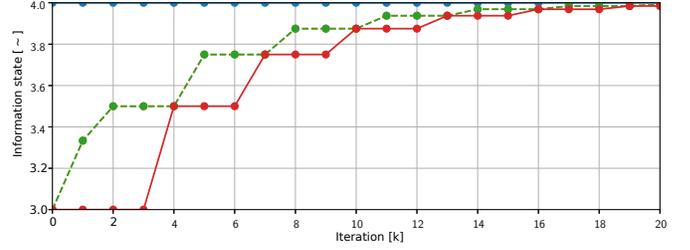}
	\caption{Evolution of agents' information states
		for Example~1. 
		The dashed lines represent $x_a(k)$ and $x_d(k)$
		(which coincide). 
		The red solid line represents the information state of agent $c$. 
		Asymptotic convergence to the max value, i.e., $x^*=x_b(0)$, can be observed.}
	\label{fig:plot_as_conv}
\end{figure}
\begin{figure}[b]
	\centering
	\includegraphics[width=\columnwidth]{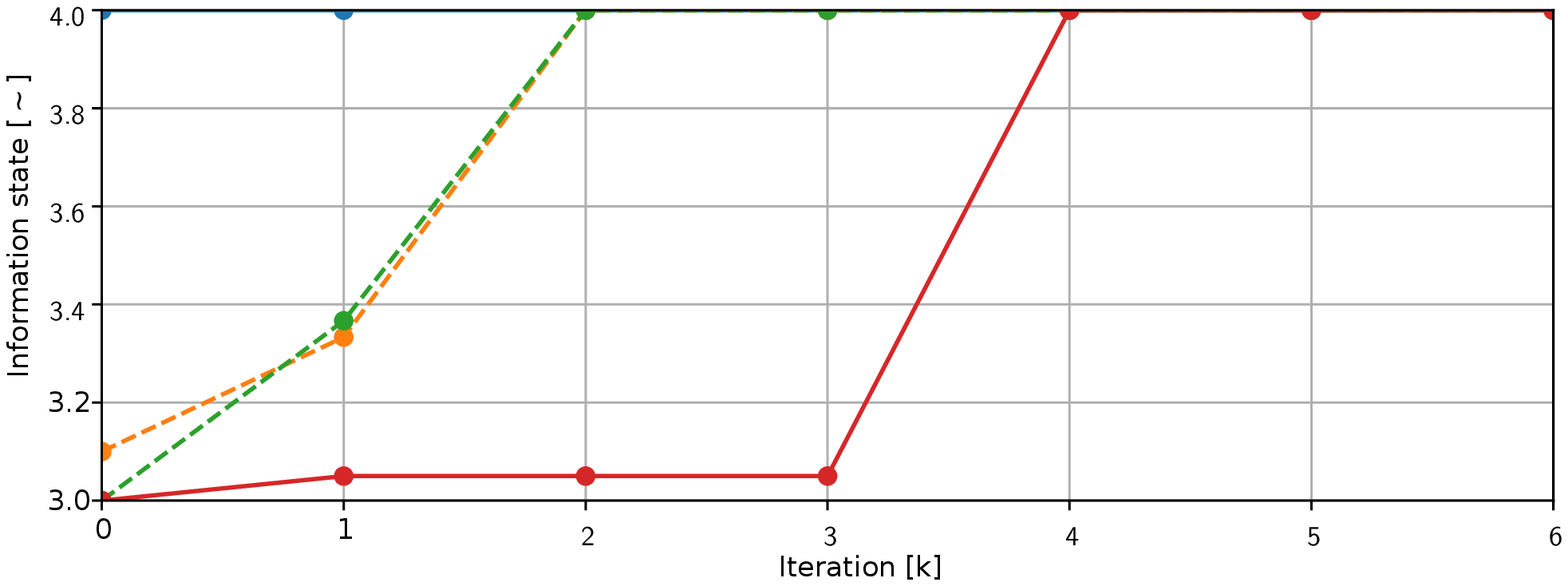}
	\caption{Evolution of agents' information states for Example 2. 
		The dashed lines represent the information states of agents $a$ and $d$, the solid
		line that of agent $c$. Max-consensus
		$x_b(0)$ is achieved after $4$ iterations.}
	\label{fig:plot_conv_x0}
\end{figure}
\begin{corollary}
	\label{prop:switchingAlwaysOff}
	\com{Consider} a multi-agent system  with a 
	\com{strongly} connected 
	network topology $(\mathcal{N},\mathcal{A})$ 
	iterating consensus protocol (\ref{eq:consMatForm}).
%	Given an arbitrary $k_0\in\natn$,
	\begin{multline}
		\com{
	%		\forall\mathbf{x}(k_0)\in\mathrm{S}^{n},\ 
			\forall i\in\mathcal{N},~%\ \forall k\geq k_0, \\
			x_i(k_0)<x^*\implies \exists k_i>k_0:\ y_i(k_i)=0.			\label{eq:eachAgentIsOvertaken}
		}
	\end{multline}
	\begin{proof}
		\com{
			Let $\ x^* - x_i(k_0)=\epsilon_0$.
			Choose $\epsilon<\epsilon_0$.
			Asymptotic stability implies that
			$\exists k_\epsilon>k_0$
			such that $x^*-x_i(k)<\epsilon$, $\forall k>k_\epsilon$.
			Hence,
			$x_i(k)>x_i(k_0)$, $\forall k>k_\epsilon$.
			This implies that $\exists k_i$,
			$k_0<k_i\leq k_\epsilon$ s.t.
			$$
				x_i(k_i)-x_i(k_i-1)>0,
			$$
			therefore $y_i(k_i)=0$.
		}
%		By Theorem \ref{prop:Lyapunov}, the system converges asymptotically to max-consensus.
%		That is to say that, given an arbitrary $\epsilon\in\mathbb{R}_{>0}$,
%		\begin{equation}
%			\label{eq:epsilon_conv}
%			\forall i\in\mathcal{N},\ %\forall \epsilon\in\mathbb{R}_{>0},\
%			\exists k_\epsilon\in\natn: %\\
%			\forall k'>k_\epsilon:\ x^* - x_i(k')<\epsilon.
%		\end{equation}
%		Let's arbitrarily select $i\in\mathcal{N}$ and $\kinN$, such that 
%		$x^*-x_i(k)=\epsilon_0$, 
%		with $\epsilon_0>0$
%		(therefore, $x_i(k)<x^*$).		
%		By (\ref{eq:epsilon_conv})
%		and by arbitrarily choosing $\epsilon<{\epsilon_0}$ ($\epsilon>0$),
%		$\exists k_\epsilon>k:\ x^*-x_i(k_\epsilon)<\epsilon$.
%		It is trivial to show that $x_i(k_\epsilon)>x_i(k)$.
%		As a consequence, 
%		$$\exists k_i\in[k+1,k_\epsilon]:\ x_i(k_i)>x_i(k_i-1),$$
%		which, by (\ref{eq:lyap_y_impl}), yields
%		$y_i(k_i)=0$.
%		This concludes the proof.
	\end{proof}
\end{corollary}
\com{The interpretation of this corollary is
that an agent
which is non-maximal at iteration $k_0$,
will eventually  lose its status as a maximal candidate
and therefore its authorization to broadcast.

In conclusion, we have shown that a
}
%
%The value of $k_i$ will intuitively depend on the network topology, 
%on the initial information state, 
%and on the realizations of the CIR.
%We will not be interested in upper bounding its value;
%for the purpose of the upcoming analysis, the existence of such a finite $k_i$
%%within the natural numbers set 
%is enough.
%Any 
multi-agent system with a \com{strongly} 
connected network topology
achieves max-consensus asymptotically by employing protocol (\ref{eq:consMatForm}).
This protocol harnesses the 
interference property of the 
channel for computing the signal $u_i$.
\com{
	In the following, we show simulation studies 
	that provide further information
	on convergence and computational benefits of the suggested
	approach.
}

\subsection{Simulations}
\begin{figure}[t]
	\centering
	\includegraphics[width=\columnwidth]{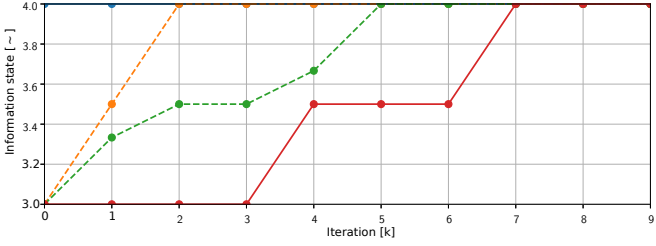}
	\caption{Evolution of agents' information states 
		for Example 3. 
		The dashed lines represent the information states of agents
		$a$ and $d$,
		the solid line that of agent $c$.
	Max-consensus
	$x_b(0)$ is achieved after $7$ iterations.}
	\label{fig:plot_conv_top}
\end{figure}

\begin{figure}[b]
	\centering
	\includegraphics[width=\columnwidth]{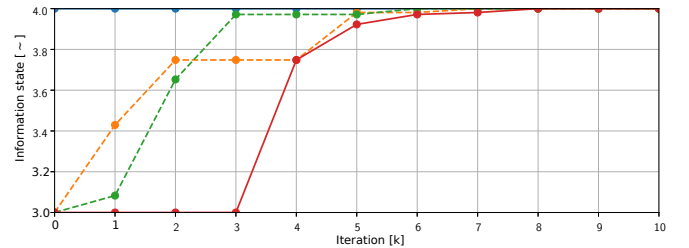}
	\caption{Evolution of agents' information states in the case of a fading wireless channel. %is considered. 
		This has an important impact on the convergence:
		with regards to Figure \ref{fig:plot_as_conv} (nonfading channel), convergence is here achieved in a finite number of steps,
		rather than asymptotically.}
	\label{fig:plot_conv_chCoeff}
\end{figure}

In the following, some simulation results are presented.
%\fa{
	The aim is to compare our approach to
	standard approaches and quantify
	benefits (saved wireless resources).
	However, as it will be clear in the simulations,
	there are cases in which finite-time convergence cannot
	be achieved with protocol (\ref{eq:consMatForm}), but only
	asymptotic convergence.
	In such cases, we cannot quantify benefits.
	This will motivate the search for an extended protocol
	in the next section.
%}
\newline
\textbf{Example 1}: first, we review a numerical
experiment in 
\cite{molinari2018exploitingMax}, where channel coefficients are equal and constant.
Consider the underlying network in Figure~\ref{fig:graph_as_conv}-\ref{fig:graph_as_conv_topDiff}.
By (\ref{eq:u_i}), $u_i(k)$ is the linear average of information states of 
agents in $N_i^m(k)$. 
Example 1 illustrates
asymptotic convergence of a system composed of 4 nodes, with a 
strongly connected network topology,
endowed with protocol (\ref{eq:consMatForm}),
see Figure~\ref{fig:plot_as_conv}.
\newline
\textbf{Examples 2-3}:
on the other hand, Examples 2, respectively Example~3,
show that, by slightly varying $\mathbf{x}(0)$, respectively the network topology,
the system achieves finite-time max-consensus
(see Figures \ref{fig:plot_conv_x0} and \ref{fig:plot_conv_top}).
This behavior has been confirmed by running extensive numerical simulations: in most cases,
finite time convergence, rather than asymptotic convergence, is achieved. 
\newline
\textbf{Examples 4}: (see Figure~\ref{fig:plot_conv_chCoeff}),
 channel coefficients are randomly drawn
out of a Rayleigh distribution with variance~1
%as in (\ref{Eq:normalizedFadChCoeff}),
%where channel impulse responses are drawn out of a Rayleigh
%distribution with variance $1$,
%(cf. Example 4),
independently for every iteration. 
Numerical experiments indicate that
the system is \textit{very likely} to achieve finite-time max-consensus. 
However, %in the case we refer to the situation analyzed in Example 1, 
it will be possible to choose a collection of constant channel coefficients,
i.e. $\forall \kinN,\ h_{ji}(k)=h$,
so that consensus is achieved asymptotically, rather than finite-time.
%%Yet, extensive numerical simulations show that, with condition \ref{eq:chCoeffStochReal}, is always achieved finite-time.
%In fact, although
%asymptotic convergence to the max-consensus is achieved deterministically (by Proposition \ref{prop:Lyapunov}),
%we cannot state the same result for finite-time consensus.
%As in \cite{iutzeler2012analysis}, 
%we could prove that, under some conditions, the convergence time is finite with probability one.
%However, this convergence property will be depending on three ingredients: 
%the network topology, 
%the initial information states vector,
%and the realization of channel coefficients.
%
%The purpose of the current work is to provide the reader with an algorithm 
%that deterministically achieves max-consensus in a finite number of steps,
%independently of these ingredients (under the trivial assumption of a connected network topology).
%To this end, consensus protocol (\ref{eq:consMatForm}) is extended 
%and the resulting protocol is presented in the next section.
\newline
These examples
illustrate that the use of protocol~(\ref{eq:consProtAsympt})
for a strongly connected network does not guarantee
finite-time consensus.
Asymptotic convergence is guaranteed by Theorem~\ref{prop:Lyapunov};
the achievement of finite-time consensus,
on the other hand, depends on the network topology,
the initial information states, and the channel coefficients
if protocol~(\ref{eq:consProtAsympt}) is used.
This is the motivation for establishing 
an extended max-consensus protocol
in the next section.

\section{Finite-time Max-Consensus Protocol}
\label{sec:FTC}
In this section,
an extended max-consensus protocol
is presented.
%It
%%harnesses the channel superposition. 
%%This protocol 
%allows every multi-agent system
%with a connected network topology 
%to achieve finite-time max-consensus
%by exploiting the channel superposition property.
%The intuition that allows for extending consensus protocol (\ref{eq:consProtAsympt})
%follows directly from Corollary \ref{prop:switchingAlwaysOff}.
%First, in the next section, the idea behind this extended consensus protocol is analyzed; 
%after that, the consensus protocol is presented and, finally, a finite-time convergence proof is proposed.
\com{
	It also exploits the channel superposition
	property. 
	However, in contrast to the protocol (\ref{eq:consMatForm}),
	it guarantees finite-time convergence for strongly connected
	network topologies.}%
\subsection{Key idea}
\label{subsec:idea_for_FTC}
Let, $\forall \kinN$,
$\mathcal{M}_k$ be the set of maximal-agents at iteration $k$, i.e.
%\begin{equation}
%	\label{eq:maxSetAgents}
	$\mathcal{M}_k=\{i\in\mathcal{N}\mid x_i(k)=x^* \}$.
%\end{equation}
%The following proposition will be of use later; it states that
%the existence of a non-maximal agent implies 
%the existence of a non-maximal agent in the neighborhood of a maximal agent:
The following proposition states that, 
at any iteration,
there exists a non-maximal agent 
if and only if there is a non-maximal agent
in whose neighborhood there is
a maximal agent.
\begin{proposition}
	\label{prop:inonmaxexistN_i}
	Given a multi-agent system  with a 
	strongly connected network topology $(\mathcal{N},\mathcal{A})$,
	then
%	and endowed with the consensus protocol (\ref{eq:consMatForm}), 
%	given an arbitrary $k_0\in\natn$
%	and $\forall \mathbf{x}(k_0)\in\mathrm{S}^n$,
%	the following holds:
	$$\exists j\in\mathcal{N}\setminus\mathcal{M}_{k}
	\iff
	\exists i\in\mathcal{N}\setminus\mathcal{M}_{k} ~\mathrm{such~that}~ N_i\cap \mathcal{M}_{k}\not=\emptyset$$
	\begin{proof}$ $
%		At arbitrary iteration $k_0$, let $j\in\mathcal{N}$ be non-maximal.
%		The necessary and sufficient proofs are analyzed separately:
		\begin{itemize}
			\item[$\impliedby$] Trivial.
			\item[$\implies$] 
			Partition $\mathcal{N}$ as $\mathcal{M}_{k}$ and 
			$\mathcal{N}\setminus\mathcal{M}_{k}$.
			Choose $j\in\mathcal{N}\setminus\mathcal{M}_{k}$ and
			$\bar{j}\in\mathcal{M}_{k}$. 
			Because of strong connectedness, 
			there is a path from
			$\bar{j}$
			to $j$.
			Clearly,
			there is at least
			one arc in this path,
			say $(\bar{i},i)$,
			such that $\bar{i}\in\mathcal{M}_{k}$
			and $i\in\mathcal{N}\setminus\mathcal{M}_{k}$. 
			As $\bar{i}$ is a neighbour of $i$,
			then
			$\bar{i}\in N_i\cap\mathcal{M}_{k}$.
%			The network topology is connected. 
%			By definition, given an arbitrary pair $\bar{i},j\in\mathcal{N}$
%			($j\not=\bar{i}$),
%			there exists a path $\mathbf{p}=\{j,j_1,\dots,j_{n_i},\bar{i}\}$ between $j$ and $\bar{i}$.
%			If $\bar{i}\in\mathcal{M}_{k_0}$ and ${j}\not\in\mathcal{M}_{k_0}$,
%			trivially, 
%			$\exists i,i^*\in\mathbf{p}$, $(i,i^*)\in\mathcal{A}$,
%			such that 
%			$i\not\in\mathcal{M}_{k_0}$,
%			$i^*\in\mathcal{M}_{k_0}$.
			%and $i^*\in N_i$.
		\end{itemize}
%		This concludes the proof.
	\end{proof}
\end{proposition}
%By Corollary \ref{prop:switchingAlwaysOff}, given an arbitrary $k_0\in\natn$
%and an arbitrary $\mathbf{x}(k_0)\in\mathbb{R}_{\geq0}^n$,
%there will exist a time-step 
%$\bar{k}>k_0$
%such that 
%$\forall i\in\mathcal{N}\setminus\mathcal{M}_{k_0},\ \exists k_i\in[k_0,\tilde{k}]:\ y_i(k_i)=0$.
%The latter leads to the following result.
The following result is derived directly from Corollary \ref{prop:switchingAlwaysOff}.
\begin{proposition}
	\label{prop:donotBroadcastIfLastInterval}
	Given a multi-agent system with a 
	strongly connected network topology $(\mathcal{N},\mathcal{A})$
	endowed with the consensus protocol (\ref{eq:consMatForm}), 
	given an arbitrary $k_0\in\natn$
	and $\forall \mathbf{x}(k_0)\in\mathrm{S}^n$,
	the following holds:
	\begin{equation}
		\label{eq:intuition_for_Extended_protocol}
		\exists \tilde{k}>k_0:\ 
		\forall i\in\mathcal{N}\setminus\mathcal{M}_{k_0},\ 
		\prod_{t=k_0}^{\tilde{k}} y_i(t)=0.
	\end{equation}
	\begin{proof}
		Corollary~\ref{prop:switchingAlwaysOff}
		states that for each $i\in\mathcal{N}\setminus\mathcal{M}_{k_0}$,
		there exists $k_i>k_0$ such that
		$y_i(k_i)=0$.
		Take $$\tilde{k}:=\max_{i\in \mathcal{N}\setminus\mathcal{M}_{k_0}} k_i.$$
		Then, $\forall i\in\mathcal{N}\setminus\mathcal{M}_{k_0}$,
		$$
			\prod_{t=k_0}^{\tilde{k}}
			y_i(t)=0.
		$$
		The proof is concluded.
	\end{proof}
\end{proposition}
By Proposition \ref{prop:donotBroadcastIfLastInterval}, each agent of the system, say agent $i$, that at $k_0$ is not maximal (i.e. $x_i(k_0)<x^*$), 
within $(\tilde{k}-k_0)$ steps will receive an input $u_i(k)$, $k\in[k_0,\tilde{k}-1]$, such that $u_i(k)>x_i(k)$.
%\begin{remark}
%	\label{rmrk:ktildedepends}
%	The value $\tilde{k}\in\natn$ depends only on the network topology $(\mathcal{N},\mathcal{A})$.
%\end{remark}
\newline
Now suppose that we change the consensus protocol
(\ref{eq:consProtAsympt})
by setting the authorization variable of each agent
$i\in\mathcal{N}$
at iteration $\tilde{k}+1$ to 
%only at iteration $\tilde{k}+1$,
%the authorization variable of each agent $i\in\mathcal{N}$ is forced to be
\begin{equation}
	\label{eq:newUpdateY}
%	\forall i\in\mathcal{N},\
	y_i(\tilde{k}+1)=\prod_{t=k_0}^{\tilde{k}} y_i(t).
\end{equation}
%(instead of (\ref{eq:updateY})),
%this would clearly imply that
%\begin{equation}
%	\label{eq:necGlob}
%	\forall i\in\mathcal{N},\ N_i\cap \mathcal{M}_{k_0}\not=\emptyset \implies 
%	x_i(\tilde{k}+2)=x^*.
%\end{equation}
%In fact, by Proposition \ref{prop:donotBroadcastIfLastInterval} and (\ref{eq:newUpdateY}),
%at time-step $\tilde{k}+1$ each non-maximal agent at $k=k_0$ does not broadcast its information state to the neighbors;
%therefore, 
From Proposition~\ref{prop:donotBroadcastIfLastInterval},
this quantity is zero for all agents that were non-maximal
at $k_0$,
implying that
\begin{equation}
	\label{eq:necGlob}
	\forall i\in\mathcal{N},\ N_i\cap \mathcal{M}_{k_0}\not=\emptyset \implies 
	x_i(\tilde{k}+2)=x^*.
\end{equation}
In other words,
all agents in the neighborhood of a maximal agent will become maximal at iteration $\tilde{k}+2$.

However, since agents do not have the global knowledge of the system, 
the value of $\tilde{k}$ is not known a priori.
Hence, there will be the need for each agent $i\in\mathcal{N}$ to retain a state variable, 
say $T_i:\natn\rightarrow\mathbb{N}$,
that attempts to (over-)estimate $\tilde{k}$.
By letting $T_i(k)$ grow according to a nondecreasing diverging sequence,
it will be eventually large enough to over-approximate $\tilde{k}$.

%%%%%%%%%%%%%%%%%%%%%%%%%%%%%%%%%%%%
%%%%% BACKUP SECTION FOR T_i %%%%%%%

%The initial value of $T_i$ is given and equal for all agents, that is to say
%\begin{equation}
%	\label{eq:initialT_i}
%	\forall i\in\mathcal{N},\ T_i(0)=T_0.
%\end{equation}
%%We hold $T_i(k+1)=T_i(k)$ until the discrete-time step $\kinN$ equals $k=2T_i(k)$;
%$T_i(k)$ is kept constant until the discrete-time step $\kinN$ equals $k=2T_i(k)$;
%when this occurs, the value of $T_i$ is updated
%according to the following:
%\begin{equation}
%	\label{eq:updateT_i}
%	\forall i\in\mathcal{N},\ 
%	k=2T_i(k)\implies T_i(k+1)=k.
%\end{equation}
%Intuitively, $T_i$ grows exponentially as the power of $2$,
%thus being eventually large enough to over-approximate $\tilde{k}$.
%Given the result provided by (\ref{eq:necGlob}), when $k=2T_i(k)$, 
%the authorization variable $y_i$ is updated coherently with what shown in (\ref{eq:newUpdateY}),
%i.e.
%\begin{equation}
%	\label{eq:new_y}
%	\forall i\in\mathcal{N},\ 
%	k=2T_i(k)\implies
%	y_i(k+1)= \prod_{t=T_i(k)}^{k} y_i(t).
%\end{equation}
%By employing such a strategy, all agents in set
%\begin{equation}
%	\{ i\in\mathcal{N} \mid \exists \bar{k}\in[T_i(k),2T_i(k)]:\ 
%	y_i(\bar{k})=0 \},
%\end{equation}
%will not broadcast, simultaneously, at instant $2T_i(k)+1$.

%%%%%%%%%%%%%%%%%%%%%%%%%%%%%%%%%%%%
%%%%%%%%%%%%%%%%%%%%%%%%%%%%%%%%%%%%

\subsection{Protocol Design}
The idea just presented
inspires the following switching consensus protocol
$\forall i\in\mathcal{N}$, $\forall \kinN$,
%On the intuition just presented, we base the following switching extended consensus protocol,
%which has to be computed at each agent $i\in\mathcal{N}$:
\begin{subequations}
	\label{eq:FTCprot}
	\begin{align}
		\intertext{if $k=2T_i(k)$:}\nonumber
		\begin{cases}
			\label{eq:sw_pr_nullify}
			x_i(k+1)=\max(x_i(k),u_i(k))\\
			y_i(k+1)=\prod_{t=T_i(k)}^k y_i(t)\\
			T_i(k+1)=k
		\end{cases}\numberthis
		,
		\intertext{else: }\nonumber
		\begin{cases}
			\label{eq:sw_pr_standard}
			x_i(k+1)=\max(x_i(k),u_i(k))\\
			y_i(k+1)=I_{\mathbb{R}_{\geq0}}(x_i(k)-u_i(k))\\
			T_i(k+1)=T_i(k)
		\end{cases}\numberthis,
	\end{align}
\end{subequations}
where 
$\forall i\in\mathcal{N}$,
$y_i(0)=1$, $x_i(0)=x_{i_0}$, $T_i(0)=T(0)=2$, and 
$\forall i\in\mathcal{N}$,
$\forall \kinN$,
$u_i(k)$ is computed as in (\ref{eq:u_i}), 
by exploiting the superposition property of the channel.
Protocol (\ref{eq:sw_pr_standard}) is identical to protocol (\ref{eq:consProtAsympt}),
except for the trivial presence of $T_i(k)$, which is, however, kept constant and does not affect the system behavior.
Only for iteration steps $k=2^n,\ n\in\mathbb{N}$, 
the proposed consensus protocol  switches to (\ref{eq:sw_pr_nullify}).
% thus employing the idea
%already presented in Section \ref{subsec:idea_for_FTC}.
% \subsection{Results}
\begin{proposition}
	\label{prop:howisT_I}
	%Note that, 
	$\forall i\in\mathcal{N}$, $\forall \kinN$, 
	\begin{equation}
		\label{eq:natureT_i}
		T_i(k)
		=
		2^{p(k)}\ 
		,
	\end{equation}
	where 	
	\begin{equation}
	\label{eq:exponentT_i}
		p(k):=
		\begin{cases}			
			\lceil\log_2(k)-1\rceil &\text{if }k\geq2\\
			1	&\text{else}
		\end{cases}
		.
	\end{equation}
	\begin{proof}
		Follows directly from (\ref{eq:FTCprot}).
	\end{proof}
\end{proposition}
Proposition~\ref{prop:howisT_I}
implies that $T_i(k)$, $k\in\mathbb{N}_0$,
is a non-decreasing diverging sequence. 
\begin{remark}
	The state variable $T_i(k)$ 
	is the same for all $i\in\mathcal{N}$, therefore, 
	the index $i$ can be omitted.% in those contexts where it is not explicitly required.
	%, $$\forall i\in\mathcal{N},\ \forall \kinN,\ T_i(k)=T(k).$$
\end{remark}
%\begin{proposition}
%	\label{prop:Tidiverging}
%%	A multi-agent system $(\mathcal{N},\mathcal{A})$ with a connected network topology
%%	is endowed with the switching consensus protocol (\ref{eq:FTCprot}).
%%	The following holds $\forall \mathbf{x}(0)\in\mathbb{R}_{\geq0}^n$ and
%%	for every realization of channel coefficients:
%	According to the switching consensus protocol (\ref{eq:FTCprot}),
%	$\{T(k)\}_{\kinN}$ is a non-decreasing diverging sequence.
%	\begin{proof}
%		%The proposition is proven directly by analyzing (\ref{eq:sw_pr_nullify})-(\ref{eq:sw_pr_standard}).
%%		Trivially, $\forall \kinN,\ \forall i \in\mathcal{N}$, $T_i(k)>0$.
%%		Also, $\forall \kinN,\ k\not=2^n,\ n\in\mathbb{N}$, $T_i(k+1)=T_i(k)$, 
%%		thus being non-decreasing in those time steps.
%%		Yet, $\forall k=2^n,\ n\in\mathbb{N}$, $T_i(k+1)=2T_i(k)$. 
%%		This shows that $T_i$ is kept constant, besides in those steps $k$ power of $2$, where it is doubled, 
%%		thus implying that $\{T_i(k)\}_{\kinN}=\{T(k)\}_{\kinN}$ form a non-decreasing diverging sequence.
%		It follows as a consequence of Proposition \ref{prop:howisT_I} 
%		and
%		(\ref{eq:natureT_i})-(\ref{eq:exponentT_i}).
%	\end{proof}
%\end{proposition}
\begin{corollary}
	\label{cor:tildek}
	A multi-agent system with a strongly
	connected network topology $(\mathcal{N},\mathcal{A})$
	employs switching consensus protocol (\ref{eq:FTCprot}).
	Then, $\forall \mathbf{x}(0)\in\mathrm{S}^n$, 
	\begin{equation}
		\label{eq:setMaxNonDecreasing}
		\exists k_s\in\natn:\ \forall k\geq k_s,\ T(k)\geq\tilde{k}.
	\end{equation}
	\begin{proof}
		By construction,
		$T(k)$ is a non-decreasing unbounded sequence.
%		By Proposition \ref{prop:donotBroadcastIfLastInterval}, $\tilde{k}$ exists, it is finite, and depends only on the network topology.
%		By Proposition \ref{prop:Tidiverging}, $\{T(k)\}_{\kinN}$ form a diverging sequence.
%		Eventually, $\exists k_s\in\natn$,\ $T(k_s)\geq\tilde{k}$.
%		Being $\{T(k)\}_{\kinN}$ also nondecreasing, it is straightforward to show (\ref{eq:setMaxNonDecreasing}).
%		By Remark \ref{rmrk:ktildedepends},
%		$\tilde{k}$ is specific 
%		to the particular network topology.
%		If $\tilde{k}\leq2$, the proof is trivial.
%		If $\tilde{k}>2$, then,
%		by (\ref{eq:natureT_i})-(\ref{eq:exponentT_i}),
%		$$
%			T(k)\geq\tilde{k} 
%			\iff 
%			\lceil \log_2(k) - 1 \rceil \geq \log_2(\tilde{k}).
%		$$
%		The latter is sufficiently verified for each time step $\kinN$, such that
%		\begin{equation}
%			\label{eq:whichkenough}
%			k
%			\geq
%			2^{
%				1 + \log_2(\tilde{k})
%			}
%			= 
%			2\tilde{k}=k_s
%			.
%		\end{equation}
%		Therefore, given a specific $\tilde{k}\in\natn$, 
%		there will eventually exist a time step 
%		$k_s\in\natn$,
%		such that,
%		$\forall k\geq k_s$,
%		$T(k)\geq \tilde{k}$.
%		The proof is completed.
	\end{proof}
\end{corollary}
\begin{remark}
	It is straightforward to come up with a suitable~$k_s$.
	In fact,
	$$
		T(k)\geq\tilde{k} 
		\iff 
		\lceil \log_2(k) - 1 \rceil \geq \log_2(\tilde{k}).
	$$
	The latter is true for each iteration $k\in\mathcal{N}$,
	such that
	\begin{equation}
		\label{eq:whichkenough}
		k
		\geq
		2^{
			1 + \log_2(\tilde{k})
		}
		= 
		2\tilde{k}=k_s
		.
	\end{equation}
\end{remark}

\begin{figure}[t]
	\centering
	\includegraphics[width=\columnwidth]{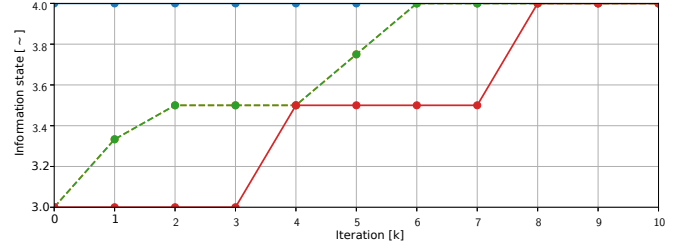}
	\caption{Evolution of information states through iterations in the absence of a fading channel.
		The solid line indicates the information state of agent
		$c$, whilst the dashed lines (overlapped)
		are the ones of agents $a$ and $d$.
		Max-consensus is achieved in $8$ iterations.
%		At discrete-time step $k=5$ (equal to $2T(0)+1$), 
%		all non-maximal agents are not allowed to broadcast; 
%		this enables the neighbors of any maximal agent ($\{a,d\}$) to achieve max-consensus
%		(at $2T(0)+2$).
%		Finally, agent $c$, get to the consensus at $k=8$.
		}
	\label{fig:plot_ft_conv_noCh}
\end{figure}

\begin{lemma}
	\label{prop:maxsetincreasing}
	A multi-agent system with a 
	strongly connected network topology $(\mathcal{N},\mathcal{A})$ 
	employs protocol (\ref{eq:FTCprot}).
	Then, $\forall \mathbf{x}(0)\in\mathrm{S}^n$, 
	\begin{equation}
		\label{eq:strictIncreasingM_theo}
		\forall k\geq k_s,\
		 \exists j\in\mathcal{N}\setminus\mathcal{M}_{T(k)}\implies \mathcal{M}_{T(k)}\subset\mathcal{M}_{2T(k)+2}.
	\end{equation}
	In words: if max-consensus has not yet been
	achieved at iteration $T(k)$,
	the number of non-maximal agents will be strictly smaller
	at iteration $2T(k)+2$.
	\begin{proof}
		%Starting from instant $k_s$, 
		%$\forall k\geq k_s$, $T_i(k)\geq\tilde{k}$.
		By Proposition \ref{prop:inonmaxexistN_i},	
		given the left-hand side of (\ref{eq:strictIncreasingM_theo}),
		there exists a maximal agent in the neighborhood of a non-maximal agent
		$i$,
		i.e. $\exists i\in\mathcal{N}\setminus\mathcal{M}_{T(k)}:
		N_i\cap \mathcal{M}_{T(k)}\not=\emptyset$.
		By %having the following protocol:
		(\ref{eq:sw_pr_nullify}),
		for $k=2T_i(k)$,
		$$\forall i\in\mathcal{N},\ y_i(2T_i(k)+1)=\prod_{t=T_i(k)}^{2T_i(k)}y_i(t).$$
		By Corollary \ref{cor:tildek},
		$\forall k\geq k_s$
		(i.e., $T_i(k)\geq \tilde{k}$),
		the following holds:
%		(which is contained in (\ref{eq:sw_pr_nullify}) for $T_i(k+1)=2T_i(k)$) 
%		and as a consequence of (\ref{eq:necGlob}),
%		$\forall k\geq k_s$, 
		\begin{equation}
			\label{eq:necGlob_Ks}
			\forall i\in\{
				i\in\mathcal{N}\mid
				N_i\cap \mathcal{M}_{T(k)}\not=\emptyset
			\},~
%			\forall i\in\mathcal{N},\ N_i\cap \mathcal{M}_{T(k)}\not=\emptyset\ 
			x_i(2T_i(k)+2)=x^*,
		\end{equation}
		meaning that agent $i$
		will become maximal at instant $2T(k)+2$.
%		%In fact, given the existence of a non-maximal agent, by Proposition \ref{prop:inonmaxexistN_i},	
%		In fact, if agent $i$ (neighbor of a maximal-agent) is already maximal at $T(k)=T_i(k)$, (\ref{eq:necGlob_Ks}) is trivial.
%		Yet, if it is not maximal at $T(k)$, it will necessarily become maximal at instant $k=2T(k)+2$.
%		As long as there exists at least one non-maximal agent $j\in\mathcal{N}$ at instant $k\geq k_s$, 
%		%such that agent $i$ is a neighbor of at least one maximal agent, i.e. $N_i\cap \mathcal{M}_{T_i(k)}\not=\emptyset$,
%		the maximal agents set at time $2T(k)+2$ will be strictly larger than the maximal agents set at time $T(k)$, 
%		thus yielding (\ref{eq:strictIncreasingM_theo}).
%		i.e. 
%		$\mathcal{M}_{T(k)}\subset\mathcal{M}_{2T(k)+2}$. Formally, $\forall k>k_s$, 
%		\begin{equation}
%			\label{eq:Mincreasingstrictly}
%			\exists j\in\mathcal{N}\setminus\mathcal{M}_{T(k)}
%			%\ N_i\cap \mathcal{M}_{T_i(k)}\not=\emptyset
%			\implies
%			\mathcal{M}_{T(k)}\subset\mathcal{M}_{2T(k)+2}.			
%		\end{equation}
%		By having 
%		$$\forall i\in\mathcal{N},\ \forall \kinN,\ T(k+1)=T_i(k+1)=2T_i(k),$$ 
%		(\ref{eq:strictIncreasingM_theo}) immediately follows and the proof is concluded.
	\end{proof}
\end{lemma}
Given the above results,
it is straightforward to show
that finite-time max-consensus is deterministically achieved
by employing protocol
(\ref{eq:FTCprot}).

\begin{figure}[t]
	\centering
	\includegraphics[width=\columnwidth]{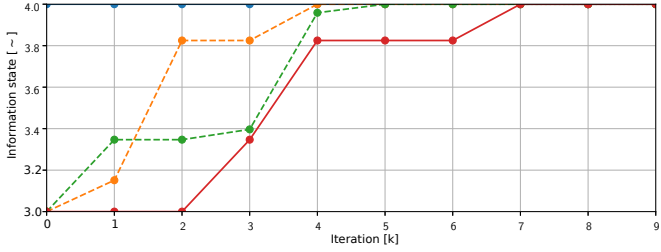}
	\caption{		
		Similar to Figure \ref{fig:plot_ft_conv_noCh}, but in the presence of a fading channel, the system achieves finite-time max-consensus.
	However, the evolution through iterations of the information states
	of agents $a$ and $c$ (dashed lines) are different. 
	This is since, in general, due to the presence of channel coefficients, $u_a(k)\not= u_c(k)$.
	}
	\label{fig:plot_ft_conv_Ch}
\end{figure}

\begin{theorem}
	Given a multi-agent system with a 
	strongly connected network topology $(\mathcal{N},\mathcal{A})$ 
	and initial information state $\mathbf{x}(0)\in\mathrm{S}^n$. 
	If agents employ the switching consensus protocol 
	(\ref{eq:FTCprot}),
	the system achieves 
	finite-time max-consensus.
	\begin{proof}
		By Lemma \ref{prop:maxsetincreasing}, 
		$\forall k\geq k_s$, 
		%every $T_i(k)+2$ steps,
		the number of maximal agents strictly increases between $k=T(k)$ and $k=2T(k)+2$,
		unless $\mathcal{N}=\mathcal{M}_{T(k)}$.
		Therefore,
		$\forall k\geq k_s$,
		\begin{equation}
			\label{eq:subsetMchain}
			\mathcal{M}_{T(k)}\subset\mathcal{M}_{2T(k)+2}\subset\dots\subseteq\mathcal{N}
			,
		\end{equation}
		which is equivalent to 
		\begin{equation}
			\label{eq:increasingMcard}
			|\mathcal{M}_{T(k)}|<|{M}_{2T(k)+2}|
			%<|\mathcal{M}_{T_i(k+2)+2}|
			<\dots\leq|\mathcal{N}|
			.
		\end{equation}
		As $\mathcal{N}$ is a finite set,
		it is obvious that this process is finished
		after a finite numbers of steps.
%		Being the cardinality of $\mathcal{N}$ finite, i.e. $|\mathcal{N}|\in\mathbb{N}$,
%		it is trivial to prove that 
%%		\begin{equation}
%%			\label{existenceKEnd}
%			$\exists \bar{\bar{k}}\in\natn :\
%			|\mathcal{M}_{2T(\bar{\bar{k}})+2}|=|\mathcal{N}|$,
%%		\end{equation}
%		which is equivalent to (\ref{eq:def_finit_time_mC}), thus showing that 
%		finite-time max-consensus is achieved at instant $2T(\bar{\bar{k}})+2 \in\mathbb{N}$.
	\end{proof}
\end{theorem}

\subsection{Simulations}
%In Example \ref{exmp:same4NodesFT}, the system $(\mathcal{N},\mathcal{A})$ (with network topology as in Figure \ref{fig:graph_as_conv})
%is endowed with the switching consensus protocol (\ref{eq:FTCprot}).
%On the other hand, Example \ref{exmp:largeNetwork} 
%studies the behavior of a larger system endowed with the switching consensus protocol (\ref{eq:FTCprot}).
%Finite-time max-consensus is always achieved, as just proven.
The following numerical experiments 
illustrates that multi-agent systems with strongly connected network topologies indeed
achieve finite-time max-consensus by employing the switching extended consensus protocol (\ref{eq:FTCprot}).
\setcounter{example}{4}
\begin{example}
	\label{exmp:same4NodesFT}
	The multi-agent system, with network topology $(\mathcal{N},\mathcal{A})$ as in Figure \ref{fig:graph_as_conv}
	and with identical and constant channel coefficients,
	is endowed with the switching consensus protocol (\ref{eq:FTCprot})
	and the simulation result is shown in Figure \ref{fig:plot_ft_conv_noCh}.
	Unlike Example~1, finite-time max-consensus is achieved.
	At instant $k=2T(0)+1=5$, all non-maximal 
	agents 
	have lost authorization to broadcast;
%	are switched off; 
	by this, at instant $k=2T(0)+2=6$,
	all those agents including
%	 the neighborhood of 
	a maximal agent in their neighbourhood
	become maximal as well.
	In the case of a fading channel,
	where
	{normalized channel coefficients are} 
	as in (\ref{Eq:normalizedFadChCoeff}),
	with channel coefficients drawn out of a Rayleigh
	distribution with variance $1$,
	finite-time max-consensus is also achieved, as
	shown in Figure \ref{fig:plot_ft_conv_Ch}.
\end{example}

\begin{figure}[h]
	\centering
	\includegraphics[width=\columnwidth]{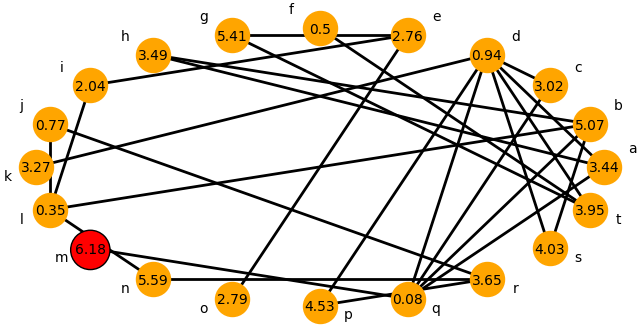}
	\caption{Multi-agent system in Example \ref{exmp:largeNetwork}
		with strongly connected 
		network topology. All arcs are directed, although (for clarity)
		directions (arrows) are omitted. In fact,
		we assume that for each arc from node $i$ to node $j$,
		there is also one arc from node $j$ to node $i$. The maximal node is $m$, and $x_m(0)=6.18$.}
	\label{fig:graph_largeNetwork}
\end{figure}

\begin{figure}[h]
	\centering
	\includegraphics[width=\columnwidth]{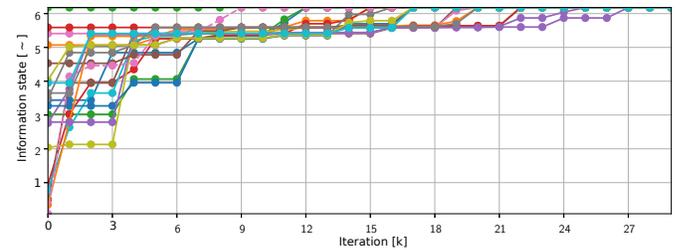}
	\caption{Evolution
	of information states for Example~6. Finite-time max-consensus is achieved in 
	$27$ steps.}
	\label{fig:plot_ft_largeNetwork}
\end{figure}

\begin{example}
	\label{exmp:largeNetwork}
	In this example, a larger system is analyzed. 
	The number of agents, the network topology,
	the channel coefficients, and the initial information state
	are randomly chosen (under the only constraint that the network topology has to be
	strongly connected),
	as shown in Figure \ref{fig:graph_largeNetwork}.
	Such a system, employed with the switching consensus protocol (\ref{eq:FTCprot}),
	achieves finite-time max-consensus, as indicated
	in Figure \ref{fig:plot_ft_largeNetwork}.
\end{example}

\subsection{Comparison with the standard approach}
In Section \ref{sec:standard_approach},
we presented the so-called standard approach.
It consists of the combination of an orthogonal
channel access communication method 
and the consensus protocol (\ref{eq:traditionalMaxCons}).
We now compare the
standard approach with the extended protocol (\ref{eq:FTCprot}),
to investigate the benefits of the latter.
%under both a convergence speed and an energy-related points of view.
\begin{figure}[b]
	\centering
	\includegraphics[width=\columnwidth]{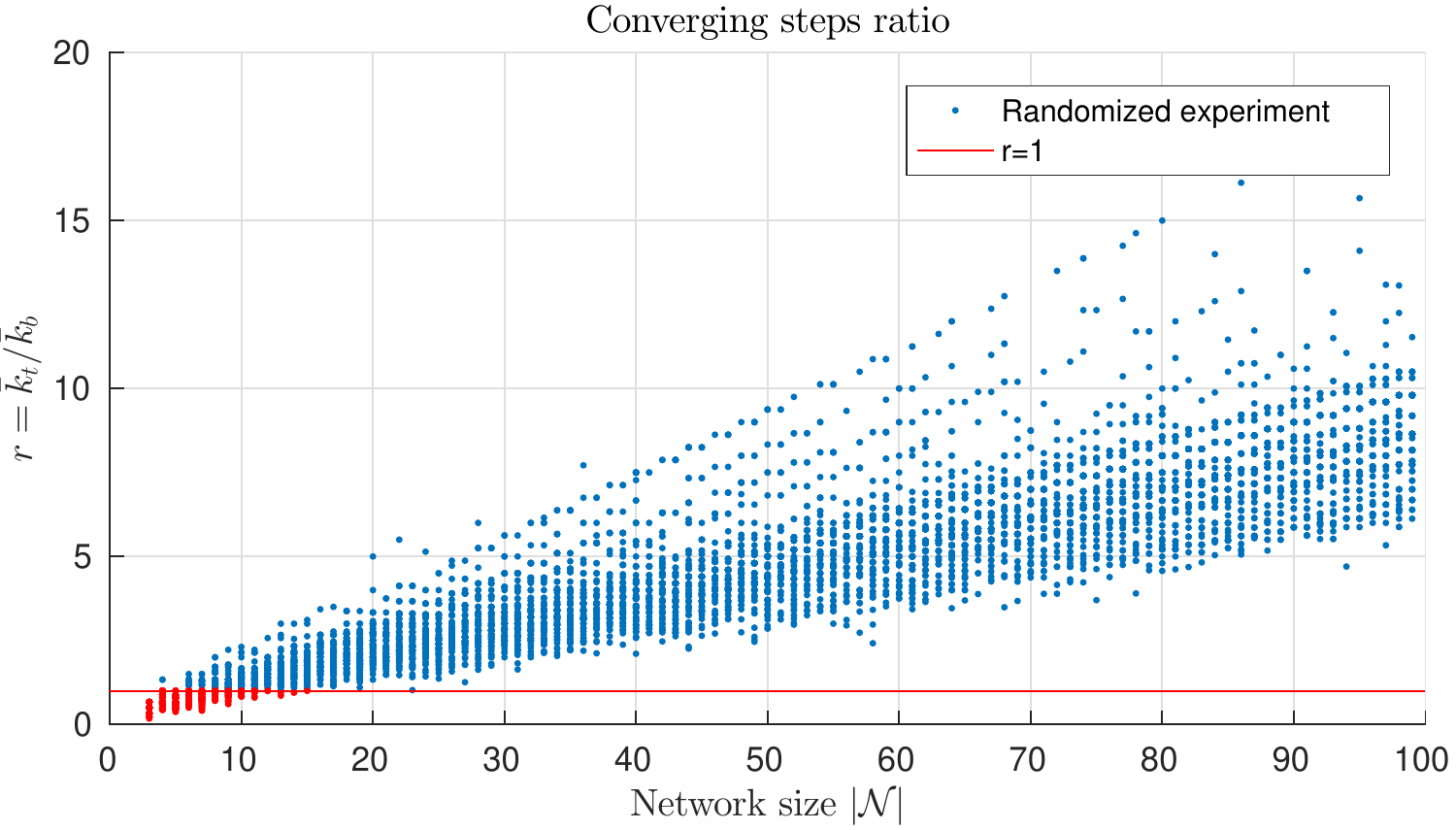}
	\caption{Each point comes from
		a randomized experiment, whose abscissa
		represents the network 
		size and its
		whose ordinate is the ratio $\bar{k}_t/\bar{k}_b$.
		All points above the red line
		correspond to those cases
		when the here proposed
		max-consensus protocol
		performs better than
		the traditional approach.}
	\label{fig:comparisonTDMA}
\end{figure}

In the following, the channel access method used for comparison is TDMA (Time-Division Multiple Access); 
this method guarantees orthogonal transmissions
by dividing each discrete transmission
into different time slots.
%Agents will broadcast sequentially, 
%one immediately after the other, each one of them using its own preassigned time-slot.
%Some coordination is required between the agents from the beginning.
%The transmission is protected against the fading and additive noise using
%digital transmission, where transmit message is converted into a sequence of binary
%digits, and protected against error by introducing redundancy using forward-error-correction coding.
%Hence, in order to broadcast a digital message from single user, multiple sub-transmissions (say $D$) are required.
Clearly, {each iteration considered in (\ref{eq:traditionalMaxCons})
	then
corresponds in reality to $n$ 
such time slots, since
each of the $n$ users
has to transmit in a one-after-the-other fashion}.

On the other hand, computing inputs 
for the agents
via superposition (cf. (\ref{eq:u_i}))
takes $2$ communication time-slots (see Section \ref{sec:commsys}),
independently of the network size,
in order to obtain the normalized channel fading coefficients.
Yet, consensus protocol (\ref{eq:FTCprot}) requires, in general, 
a higher number of iterations than the standard approach, 
and it depends on channel realization.
Therefore, a meaningful
comparison can be only 
done via randomized simulations.
%A suitable value of $M$ can 
%be chosen as $3$, see \hl{[X, Pag. X]}.

For networks of size between $3$ and $100$, 
one
hundred different simulations are executed. 
Each one represents a random initial vector
and a random (connected) topology. 
$\forall \kinN$, 
$\forall i\in\mathcal{N},\ \forall j\in N_i(k)$,
the random channel coefficients 
$\xi_{ij}(k)$ are drawn out of independent
Rayleigh distributions with variance $1$.
$\bar{k}_t$ denotes the number
of time slots
required by the traditional approach for achieving max-consensus, 
and
$\bar{k}_b$ the number of time-slots
required by the switching protocol (\ref{eq:FTCprot}) to ensure max-consensus.
For each experiment, in Figure \ref{fig:comparisonTDMA}, we plot the ratio of the two quantities,
defined as $r=\frac{\bar{k}_t}{\bar{k}_b}$.
The numerical experiment
shows that
for multi-agent systems composed of more than approximately
 {$15$ agents}, 
employing (\ref{eq:FTCprot})
and channel superposition
saves significant convergence 
time.
% Fabio
\section{Conclusion}
\label{sec:concl}
This paper has presented
a possible solution for
achieving max-consensus
in multi-agent systems communicating over real fading 
wireless channels.
First, a suitable communication
system
has been designed.
By employing this strategy,
a max-consensus protocol
adopting broadcast authorizations
has been proven to guarantee asymptotic
convergence.
Then, this protocol
has been extended with
%thus resulting in 
a switching
protocol guaranteeing
finite-time convergence.

Future work will consider
the relaxation of some of the
assumptions made
in the paper.
In particular,
we will study 
the case of a 
noisy channel,
and we will investigate the effort of asynchronous broadcasts.

\bibliography{bibliography.bib}
\bibliographystyle{plain}
\end{document}